\documentclass[11pt,a4paper,natbib]{article}
\usepackage{jcappub}
\usepackage[usenames,dvipsnames]{xcolor}

\begin{document}

\title{Impedance Matching to Axion Dark Matter: Considerations of the Photon-Electron Interaction}

\author[a,b]{Saptarshi Chaudhuri}
\affiliation[a]{Department of Physics, Princeton University, Princeton, NJ 08544}
\affiliation[b]{Department of Physics, Stanford University, Stanford, CA 94305}
\emailAdd{sc25@princeton.edu}

\date{\today}

\abstract{
We introduce the concept of impedance matching to axion dark matter by posing the question of why axion detection is difficult, even though there is enough power in each square meter of incident dark-matter flux to energize a LED light bulb. By quantifying backreaction on the axion field, we show that a small axion-photon coupling does not by itself prevent an order-unity fraction of the dark matter from being absorbed through optimal impedance match. We further show, in contrast, that the electromagnetic charges and the self-impedance of their coupling to photons provide the principal constraint on power absorption integrated across a search band. Using the equations of axion electrodynamics, we demonstrate stringent limitations on absorbed power in linear, time-invariant, passive receivers. Our results yield fundamental constraints, arising from the photon-electron interaction, on improving integrated power absorption beyond the cavity haloscope technique. The analysis also has significant practical implications, showing apparent tension with the sensitivity projections for a number of planned axion searches. We additionally provide a basis for more accurate signal power calculations and calibration models, especially for receivers using multi-wavelength open configurations such as dish antennas and dielectric haloscopes.} 

\maketitle

\section{Introduction}
\label{sec:intro}

The QCD axion and axion-like particles are leading candidates for cold dark matter. QCD axions not only possess natural mechanisms for generating the dark matter abundance\cite{Dine:1982ah,Preskill:1982cy,abbott1983cosmological,graham2018stochastic,takahashi2018qcd}, but also solve the strong CP problem\cite{Peccei:1977hh,wilczek1978problem,weinberg1978new}. Recent theoretical investigations and the advent of precision experimental techniques have resulted in the proposal and construction of several new probes for axion dark matter\cite{graham2015experimental,sikivie2021invisible}. (We refer to the QCD axion and axion-like particles collectively as ``axions.'') Many of these probes search for the axion's coupling to two photons, quantified by coupling $g_{a\gamma \gamma}$, so that in a background electromagnetic field, the axion converts to a photon\cite{Sik83,Sik85}. The photon signal may be detected with a sensitive receiver. Over much of the allowed parameter space, axion dark matter possesses a low mass $\lesssim$1 eV, which when combined with the local dark-matter density $\sim 0.45$ GeV/cm$^{3}$\cite{tanabashi2018review}, results in large number density. In the context of detection, axions are then more appropriately described as wave-like dark matter, rather than particle-like dark matter. The photon signal is best described as a coherent electromagnetic field.

In experimental searches, the background electromagnetic field usually takes the form of a several-Tesla DC magnetic field extending over a $\sim$1 m$^{3}$ volume, and the receiver often consists of a high-Q ($\gtrsim 10^{4}$) resonator. In the presence of the DC magnetic field, a nonrelativistic axion dark-matter field of mass $m_{a}$ produces an electromagnetic signal oscillating at $\omega_{a} = m_{a}c^{2}/\hbar$. If the resonance frequency of the receiver is near the rest-mass frequency, the electromagnetic signal is enhanced. One may then conduct a sensitive search for axion dark matter over a wide range of mass-coupling parameter space by tuning the resonance frequency. However, because the axion is feebly coupled to the photon, the expected power from a QCD axion signal in a state-of-the-art resonant receiver is $\lesssim$ 10$^{-22}$ Watts\cite{brubaker2017first,du2018search,zhong2018results,braine2020extended}. The low signal power necessitates the use of layers of shielding to mitigate electromagnetic interference\cite{brubaker2018first}, as well as cryogenic operation to reduce thermal noise. Additonally, the experiments utilize sensitive readout, often in the form of amplifiers operating near the Standard Quantum Limit (SQL) on phase-insensitive amplification\cite{caves1982quantum,clerk2010introduction}. Recently, there has been work on squeezed-state receivers, single-photon counting, backaction evasion, and other protocols which evade the SQL and enhance search sensitivity\cite{lehnert2016quantumaxion,backes2021quantum,dixit2021searching,chaudhuri2019dark}. %\sch{Cite Stephen's RQU paper.} 

The power available per unit area in the axion dark-matter field is given by the incident energy flux, roughly equal to the product of the local dark-matter density\cite{tanabashi2018review} and the virial velocity $\sim 10^{-3}c$. In each square meter of flux, there is then $\sim$10 Watts of power, enough to turn on a LED light bulb and $\sim$23 orders of magnitude more power than expected in resonant searches. This observation begs the question: Why is axion dark matter detection difficult? For nearly four decades, the cavity haloscope has been the standard experimental technique for electromagnetic probes of axion dark matter. Is there a fundamental reason why a vastly better absorber has not been realized? The axion-photon coupling $g_{a \gamma \gamma}$ may be ``small", but that alone does not translate to a small power absorption. Equivalently, it alone does not prevent coupling to a matched load. (See also penultimate paragraph of Section \ref{ssec:circuits}.) So what precisely are the physical mechanisms that prevent one from absorbing all power in the axion field, or equivalently, why is it impractical to impedance match to axion dark matter, necessitating the use of precision measurement techniques that are often challenging to implement? 

In this paper, we investigate impedance matching to axion dark matter, elucidating the performance of laboratory searches through conservation of energy arguments. Since the axion mass is a priori unknown, the metric we use for describing the difficulty of obtaining an impedance match is signal power integrated over a search band of rest-mass frequencies. Our results, which extend the impulse response technique introduced in ref. \cite{lasenby2021parametrics} and put a fundamental limit on frequency-integrated signal power, for the first time demonstrate the stringent constraints on coupling power from the axion-photon interaction imposed by the photon-electron interaction. We thus explain stringent, fundamental constraints on improving power absorption beyond the cavity haloscope paradigm, informing practical enhancements in future searches. While the literature on electromagnetic searches tends to focus on the axions and photons, our impedance-matching analysis shows that, in fact, the \emph{electromagnetic charges} provide the most important constraint on frequency-integrated power absorption, significantly impacting experimental sensitivity.

%The vast majority of laboratory receivers fall into the category of linear, time-invariant, and passive. This category will thus be the focus of our work. 
The paper is organized as follows. We begin in Section \ref{sec:gedank} with a thought experiment analyzing the impedance match to axion dark matter in two conceptual receivers, a broadband free-space antenna and a cavity resonator. By using equivalent circuits and by determining the axion source impedance, we quantify the backreaction of the receivers on the dark matter field. 
The calculation shows that the self-impedance of the photon-electron coupling is the cause of large impedance mismatch with axions and poor power absorption. In Section \ref{sec:GBP}, using conservation of energy and impulse response arguments, we formalize the limitations from photon-electron coupling observed in our thought experiment. We derive an upper bound on the gain-bandwidth product of the impedance match to axion dark matter (i.e. an upper bound on frequency-integrated signal power) for linear, time-invariant, passive receivers; linear, time-invariant, passive describes the vast majority of laboratory searches. In Appendix \ref{ssec:imp_response}, we explain why the argument for a gain-bandwidth relation in ref. \cite{lasenby2021parametrics}, which introduces impulse response, is incomplete, as it does not apply to systems with electrons. Incorporating the charges is a key part of our work. In Section \ref{sec:search_opt}, we use the Bode integral theorem\cite{bode1945network,fano1950theoretical} to systematically generate large classes of receivers---including the cavity resonator, widely used in axion searches--- that saturate the gain-bandwidth bound on the match to the dark matter source. In Section \ref{sec:implications}, we discuss practical implications of our analysis of photon-electron interaction and the consequent gain-bandwidth bound. We illustrate apparent tension between our results and sensitivity projections of a number of planned searches and provide a foundation for more accurate signal power calculations and calibration models, especially for multi-wavelength open receivers such as dish antennas\cite{horns2013searching} and dielectric haloscopes\cite{caldwell2017dielectric,baryakhtar2018axion}. We conclude in Section \ref{sec:conclude}.

%------------------------------------------------------

\section{Thought Experiment}
\label{sec:gedank}

\subsection{Setup}
\label{ssec:setup}

\begin{figure}[htp] 
\includegraphics[width=\textwidth]{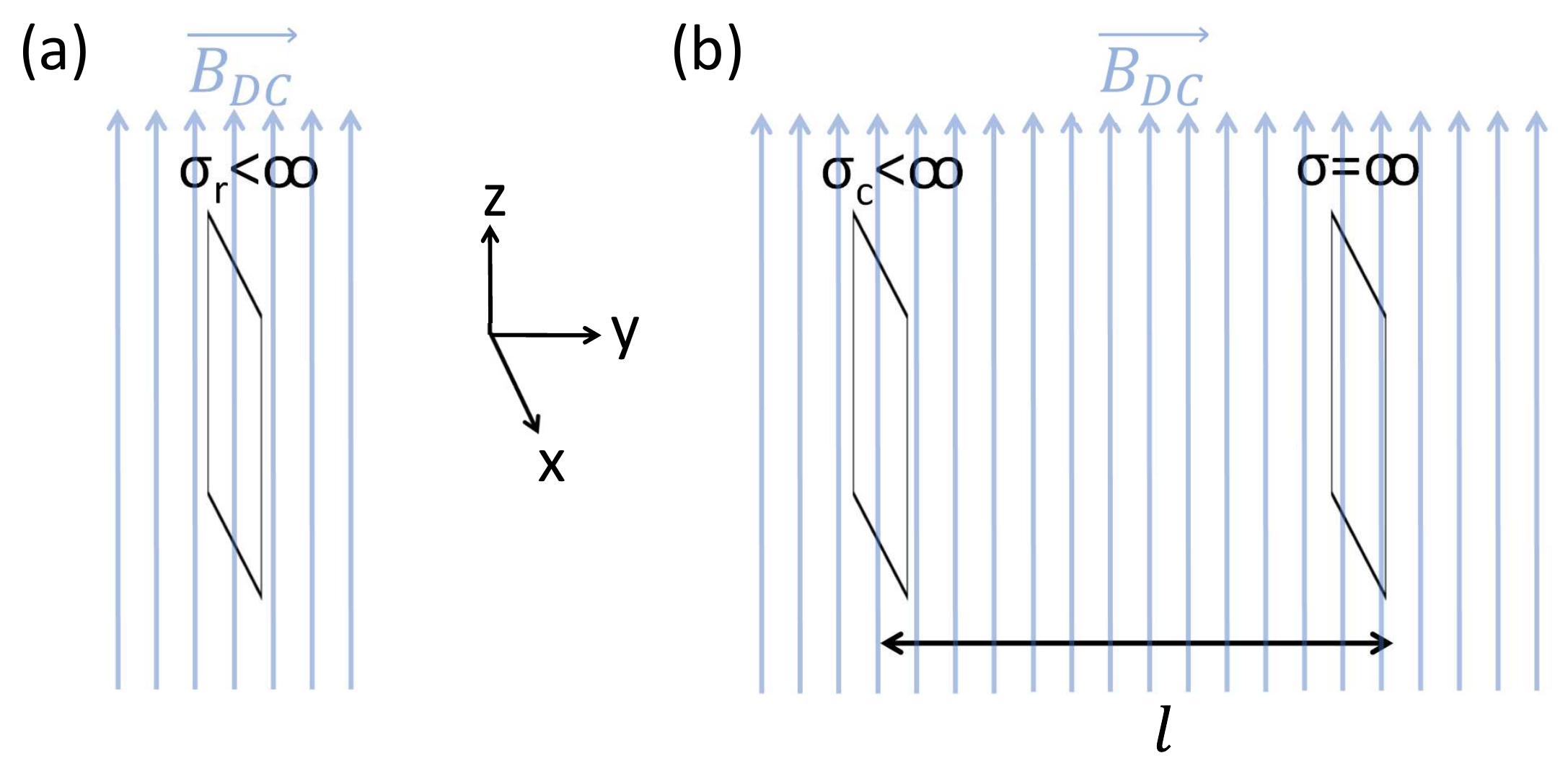}
\caption{(a) Resistive sheet antenna and (b) resonant cavity embedded in a background DC magnetic field, $\vec{B}_{DC}=B_{0}\hat{z}$. \label{fig:SheetvCavity}}
\end{figure}

Fig. \ref{fig:SheetvCavity} illustrates our thought experiment with an antenna and a cavity. As can be derived from conservation of energy\cite{chaudhuri2018fundamental,chaudhuri2019optimal,chaudhuri2019dark}, these receivers represent the two generic categories of coupling to the electromagnetic excitation from the axion: radiative coupling (antenna), through free-space radiation modes of the receiver, and reactive coupling (cavity), through energy-storing elements of the receiver. We describe the toy model and the power absorption from the axion driving field. Our analysis builds on a more detailed calculation in \cite{chaudhuri2018fundamental}.

Consider a uniform DC magnetic field of strength $B_{0}$ pointing in the $\hat{z}$ direction and filling all space. An approximately static, uniform axion field $a(t)$ of mass $m_{a}$ oscillating at its rest-mass frequency $\omega_{a}=m_{a}c^{2}/\hbar$,
\begin{equation}\label{eq:axion_field}
    a(t)=Re(\tilde{a}\textrm{exp}(+i\omega_{a}t)),
\end{equation}
produces an electric field,\cite{caldwell2017dielectric}
\begin{equation}\label{eq:Ea_field}
    \vec{E}_{a}(t)=Re(\tilde{E}_{a}\textrm{exp}(+i\omega_{a}t))\hat{z},
\end{equation}
where $\tilde{E}_{a}=\kappa_{a}cB_{0}\tilde{a}$ and $\kappa_{a}$ is related to the axion-photon coupling $g_{a\gamma\gamma}$ by $\kappa_{a}=g_{a\gamma\gamma}\sqrt{\hbar c \epsilon_{0}}$. The tilde above $E_{a}$ on the right-hand side of eq. (\ref{eq:Ea_field}) indicates that the quantity is a complex number. This electric field drives the receivers, depositing axion-field power.

In Fig. \ref{fig:SheetvCavity}(a), we show a square resistive sheet of area $A$ and ohmic conductivity $\sigma_{r}$. Suppose that its thickness $h$ is much less than the skin depth at frequency $\omega_{a}$ and that each lateral dimension is much larger than the Compton wavelength $\lambda_{a}=2\pi c/\omega_{a}$. 
The resistive sheet is then a toy model of a broadband-antenna search for axions\cite{horns2013searching}. The sheet currents, driven by the tangential field $\vec{E}_{a}(t)$ in the $\hat{z}$-direction, both dissipate power and produce electromagnetic radiation, which can be approximated as $\hat{z}$-polarized plane waves propagating in the $+\hat{y}$- and $-\hat{y}$-direction. These waves also represent the antenna receiving modes. One can solve Maxwell's equations for the steady-state power dissipation per unit area:
\begin{equation} \label{eq:Pdiss_sheet}
\frac{P_{r}}{A}= \frac{Z_{r}}{2} \frac{|\tilde{E}_{a}|^{2}}{(Z_{r} + Z_{fs}/2)^{2} },
\end{equation}
where $Z_{r}=1/(\sigma_{r}h)$ is the sheet impedance and $Z_{fs}= \sqrt{\mu_{0}/\epsilon_{0}} \approx 377\ \Omega$ is the free-space impedance. Note that the power dissipation (\ref{eq:Pdiss_sheet}) is similar to that for bolometric absorption of an electromagnetic plane wave\cite{hadley1947reflection,jones1953general}.

In Fig. \ref{fig:SheetvCavity}(b), we show two square sheets of area $A$ separated by length $\ell$. The sheet on the left possesses finite-conductivity $\sigma_{c}$, while that on the right is an ideal conductor. As with the antenna in Fig. \ref{fig:SheetvCavity}(a), the thickness $h$ of the finite-conductivity sheet is assumed to be much less than the skin depth, and the lateral dimensions are assumed to be much larger than the Compton wavelength. This receiver is a toy model of a cavity. For rest-mass frequencies $\omega_{a}$ near the half-wavelength resonance frequency $\omega_{r}=\pi c/\ell$, $\vec{E}_{a}(t)$ drives resonantly enhanced currents on the left-side sheet, resulting in enhanced energy storage between the sheets and enhanced power dissipation. Solving Maxwell's equations for the power dissipation per unit area gives, for frequencies $\omega_{a}$ near resonance, $|\omega_{a}-\omega_{r}| <<\omega_{r}$,
\begin{equation}\label{eq:Pdiss_cav}
    \frac{P_{c}}{A}\approx \frac{4}{\pi} \frac{|\tilde{E}_{a}|^{2}}{Z_{fs}} \frac{Q}{|1+2iQ (\omega_{a}-\omega_{r})/\omega_{r}|^{2}}
\end{equation}
where $Z_{c}=1/(\sigma_{c}h)$ is the impedance of the sheet on the left and $Q=\pi Z_{fs}/2Z_{c}$ is the mode quality factor.

For signals on resonance, taking the limit $Q \rightarrow \infty$ yields $P_{c}/A \rightarrow \infty$. Of course, extracting infinite power from the finite-power axion field is unphysical. Two assumptions break down. First, the axion dark matter has some velocity. The resulting linewidth for a virialized axion signal is $\sim 10^{-6} \omega_{a}$; for $Q$ much larger than $10^{6}$, our expression for dissipated power (\ref{eq:Pdiss_cav}) is then not valid, as portions of the signal are off resonance. Second, and more importantly for analyzing impedance matching, at sufficiently high $Q$, the cavity backreacts on the axion field, modifying the local axion-field amplitude via photon-to-axion conversion and effectively changing the value of $\tilde{E}_{a}$. Backreaction can be ignored as long as the dissipated power is much smaller than the available power, i.e. as long as the impedance match to axions is poor. The available power per unit area is the axion energy flux, roughly equal to the product of the energy density 
\begin{equation}\label{eq:rho_a}
    \rho_{a}=\frac{\epsilon_{0}}{2} \left(\frac{\omega_{a}}{c} \right)^{2} |\tilde{a}|^{2}
\end{equation}
and velocity $v$:
\begin{equation}\label{eq:P_axion}
    \frac{P_{a}}{A} \sim \frac{1}{2} \frac{( \omega_{a}/c)^{2} |\tilde{a}|^{2}}{Z_{fs}} \frac{v}{c}.
\end{equation}
From (\ref{eq:Ea_field}) and (\ref{eq:Pdiss_cav}), backreaction may then be ignored if
\begin{equation}\label{eq:Q_axion}
    Q \ll \left( \frac{\kappa_{a}c B_{0}}{\omega_{a}/c} \right)^{-2} \frac{v}{c}.
\end{equation}
For virialized DFSZ axions\cite{dine1981simple} in a $B_{0}=10$ T magnetic field, the quantity in parentheses is $\approx3 \times 10^{-16}$, so we may ignore backreaction as long as $Q\ll 10^{28}$. This is readily satisfied for practical cavities. Similarly, we may ignore backreaction for the antenna because the dissipated power is a small fraction, $\lesssim 10^{-28}$, of the available power for all sheet impedances. %\sch{Check numbers with Peter.}

\subsection{Equivalent Circuits}
\label{ssec:circuits}

We conclude that both receivers in the toy model are, in practice, a poor impedance match to dark matter. However, our calculations have provided no physical intuition as to why the match is poor. As Schwinger and Dicke point out\cite{schwinger1968discontinuities,schwinger1969julian,milton2007appreciation, montgomery1987principles}, equivalent circuits are powerful tools for understanding the physics of electromagnetic systems. The equivalent circuits of the receivers may readily be derived using standard techniques\cite{chu1948physical,zmuidzinas2003thermal,montgomery1987principles}.

\begin{figure}[htp] 
\includegraphics[width=\textwidth]{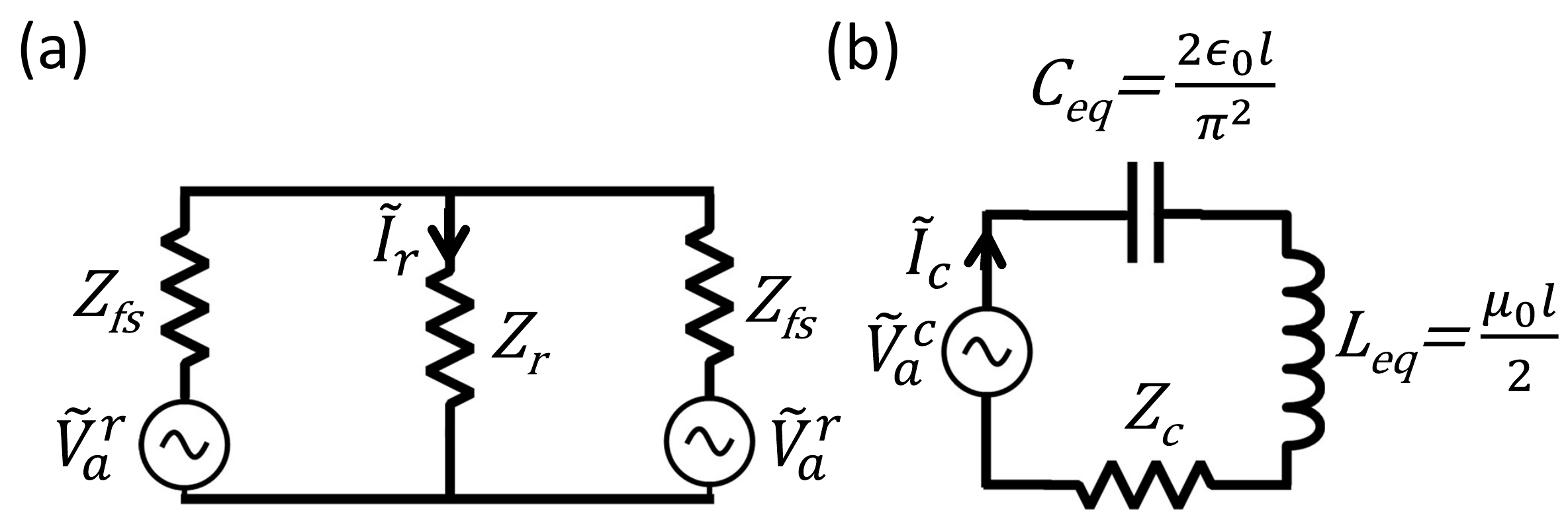}
\caption{Equivalent circuits for the (a) sheet antenna and (b) half-wave cavity mode. Axion source impedance not shown. \label{fig:CircuitModels}}
\end{figure}

In Fig. \ref{fig:CircuitModels}(a), the sheet of impedance $Z_{r}$ has been replaced by an equivalent resistor $Z_{r}$. The two radiation modes are represented by resistors of value equal to the free-space impedance $Z_{fs}$. The electric field $\vec{E}_{a}(t)$ driving the antenna is represented by two voltages---one for each receiving mode---with complex-valued voltage phasor $\tilde{V}_{a}^{r}$. As can be derived from Kirchhoff's Laws, under the mapping $\tilde{V}_{a}^{r} \rightarrow \tilde{E}_{a} \sqrt{A}$, the power dissipation in resistor $Z_{r}$ from driven current $\tilde{I}_{r}$ matches eq. (\ref{eq:Pdiss_sheet}).

%The voltage drive produces a current $\tilde{I}_{r}$ through the resistor $Z_{r}$, which using Kirchhoff's Laws, results in power dissipation
%\begin{equation}\label{eq:Pdiss_sheet_circuit}
%    P_{r}^{\rm equiv}= \frac{Z_{r}}{2}|\tilde{I}_{r}|^{2}= \frac{Z_{r}}{2} \frac{|\tilde{V}_{a}^{r}|^{2}}{(Z_{r}+Z_{fs}/2)^{2}}.
%\end{equation}
%\begin{equation}
%    \tilde{I}_{r}= \frac{2\tilde{V}_{a}^{r}}{2Z_{r}+Z_{fs}}.
%\end{equation}
%The power dissipation in the resistor $Z_{r}$ is then
%Mapping $\tilde{V}_{a}^{r} \rightarrow \tilde{E}_{a} \sqrt{A}$, we recover eq. (\ref{eq:Pdiss_sheet}).

In Fig. \ref{fig:CircuitModels}(b), the half-wave cavity mode has been represented as an equivalent series-RLC circuit. The equivalent inductor $L_{eq}$ and capacitor $C_{eq}$ represent electromagnetic energy storage between the two sheets. The sheet of impedance $Z_{c}$ is represented by a resistor of the same value. The forcing from axions is represented as a voltage $\tilde{V}_{a}^{c}$ and produces current $\tilde{I}_{c}$. Mapping $\tilde{V}_{a}^{c} \rightarrow 2 \tilde{E}_{a} \sqrt{A}$, the power dissipation in resistor $Z_{c}$ matches eq. (\ref{eq:Pdiss_cav}).

%For an axion drive (\ref{eq:axion_field}) of frequency $\omega_{a}$, the current phasor is related to the voltage phasor by
%\begin{equation}\label{eq:I_cav_approx}
%    \tilde{I}_{c}= \frac{\tilde{V}_{a}^{c}}{Z_{c}+i\omega_{a}L_{eq} + 1/(i\omega_{a}C_{eq})}\approx  \frac{\tilde{V}_{a}^{c}}{Z_{c}+2i(\omega_{a}-\omega_{r})L_{eq}},
%\end{equation}
%The power dissipation in the equivalent resistor $Z_{c}$ is
%\begin{equation}\label{eq:Pdiss_cav_circuit}
%    P_{c}^{\rm equiv}=\frac{Z_{c}}{2} |\tilde{I}_{c}|^{2} \approx \frac{Z_{c}}{2} \frac{|\tilde{V}_{a}^{c}|^{2}}{|Z_{c}+2i(\omega_{a}-\omega_{r})L_{eq}|^{2}}
%\end{equation}
%where the approximation holds for frequencies $\omega_{a}$ near resonance $\omega_{r}= 1/\sqrt{L_{eq}C_{eq}}$. Using $Q=\omega_{r}L_{eq}/Z_{c}$ and mapping $\tilde{V}_{a}^{c} \rightarrow 2 \tilde{E}_{a} \sqrt{A}$, we recover eq. (\ref{eq:Pdiss_cav}). %The factor of 2 in $\tilde{V}_{a}^{c}$, relative to $\tilde{V}_{a}^{r}$, is the result of the field $\vec{E}_{a}(t)$ coherently driving the two sheets of the cavity, as opposed to the single sheet of the resistive antenna.

As discussed earlier, at sufficiently high $Q$, the cavity backreacts on the axion field. Backreaction can be represented with a source impedance $Z_{a}^{s}$ in series with the voltage drive. The scale of the axion source impedance is given by the impedance $Z_{c}$ at which on-resonance power dissipation is comparable to available power (\ref{eq:P_axion}), i.e. the sheet impedance at which an efficient match is obtained:
\begin{equation}\label{eq:Za}
    Z_{a}^{s} \sim \left( \frac{\kappa_{a}c B_{0}}{\omega_{a}/c} \right)^{2} Z_{fs} \frac{c}{v}.
\end{equation}
See \cite{chaudhuri2019dark} for a calculation of the source impedance, solved directly from the equations of axion electrodynamics. 

For virialized DFSZ axions in a 10 Tesla field, the source impedance is $Z_{a}^{s} \sim 10^{-28} Z_{fs}$. Comparing the source impedance to the impedance scales in the circuits now reveals the mechanism limiting an efficient match to axions in linear, time-invariant, passive receivers. 

The axion produces electromagnetic fields (e.g., $\vec{E}_{a}(t)$ in the toy model), and the receiver charges couple to these fields radiatively (as with the antenna) and/or reactively (as with the cavity)\cite{chaudhuri2018fundamental,chaudhuri2019dark}. However, for either manner of coupling, in order for the charges to couple to the fields, the charges must produce fields themselves. These fields produced by the charges exert a self-impedance on the charges that is much larger than the axion source impedance, limiting electromagnetic current flow and axion power coupled to the receiver. \emph{In a linear, time-invariant, passive receiver, the impedance match to axion dark matter is limited in large part by the self-impedance of photons acting on electromagnetic charges in the receiver.} For example, in the antenna, the charges couple to the axion-induced electric field via free-space radiation modes. The radiative coupling is accompanied by the emission of electromagnetic radiation from the sheet. The radiation self-impedance on the charges, given by $Z_{fs}$, is much larger than the axion source impedance, and the mismatch limits the power coupled into the charges. Quantitatively, the impedance mismatch can be understood from eq. (\ref{eq:Pdiss_sheet}). If we eliminate the free-space impedance term in the denominator, we may absorb arbitrarily large amounts of power by reducing the sheet impedance (up to backreaction on the axion field, characterized by source impedance $Z_{a}^{s}$). However, with the free-space self-impedance in the denominator, power coupled to the receiver is severely limited. 

For the cavity in Figs. \ref{fig:SheetvCavity}(b) and \ref{fig:CircuitModels}(b), the self-impedance is both a self-capacitance and self-inductance, representing the field energy produced by the charges and stored in the cavity. On-resonance, the impedance from the inductance and capacitance cancel. One can absorb an order-one fraction of the available power by increasing the $Q$ so that the resistance $Z_{c}$ matches the source impedance $Z_{a}^{s}$\cite{chaudhuri2019dark}. In theory, an efficient match can then be obtained regardless of the value of the axion-photon coupling. However, the $Q$ required for the match is too high to be achieved practically, as shown in eq. (\ref{eq:Q_axion}). 

Off resonance, the inductive and capacitive impedances do not cancel. 
%The magnitude of the inductive impedance near resonance is approximately $\pi Z_{fs}/2$ and similarly for the capacitive impedance. Both impedances are much larger than the axion source impedance. 
As the detuning between the axion rest-mass frequency and resonance frequency increases, the net self-reactance becomes large relative to the axion source impedance. The consequent impedance mismatch limits power coupled to the cavity. Because passive reactances are necessarily frequency-dependent (see, e.g., Foster's reactance theorem \cite{foster1924reactance}), the frequency range over which the self-impedance may be nulled for an efficient impedance match is limited. Note that, although challenging in practice, this bandwidth limitation may be alleviated by the use of active-matching elements, through which charges supply energy to the receiver. For example, negative inductors and materials with negative dispersion may be used to cancel the self-reactance of a pickup over a wide range of frequencies (as opposed to a single frequency in a single-pole resonator)\cite{sussman2009non,salit2010enhancement,shlivinski2018beyond}.\footnote{It is useful to distinguish our description of impedance matching for axion waves from descriptions for electromagnetic waves and gravitational waves in the literature. For electromagnetic waves, an efficient impedance match is practical because one can match the impedance of the absorber (containing the electrons) to the impedance of the source, given by the free-space impedance $Z_{fs}$\cite{hadley1947reflection,jones1953general}. For gravitational waves, an efficient impedance match is impractical because the impedance of the absorber (the test mass), characterized by the acoustic impedance, is far below the impedance of the source, given by the impedance of spacetime\cite{blair2005detection}. For axions, an efficient impedance match is impractical because of the intermediary (the photons) between the source (the axions) and the absorber (the electrons), which results in a mismatch due to self-impedance.}

%------------------------------------------------------

\section{Gain-Bandwidth Bounds}
\label{sec:GBP}

We now formalize the observed limitations imposed by the photon-electron interaction in the form of a gain-bandwidth bound on the impedance match to axion dark matter, constraining frequency-integrated signal power.

\subsection{Conservation of Energy}
\label{ssec:COE}

%Here, we state assumptions on our derivation of energy-conservation for an axion dark-matter field interacting with a receiver. The assumptions also underlie the gain-bandwidth bound. 

Here, we derive a conservation of energy statement for an axion dark-matter field interacting with a receiver. This statement, derived from the equations of axion electrodynamics, will be used to quantify the concept of impedance seen by the axion source in Section \ref{ssec:impedance}. We assume that axion-to-photon conversion occurs in a background DC magnetic field of finite spatial extent (unlike the infinite spatial extent of the toy model), similar to those used near-universally in laboratory dark-matter searches. However, similar treatments may be produced for other types of background fields. See Appendix \ref{sec:GBP_other}. 

We take the axion field to be spatially uniform over the extent of the DC magnetic field. This is a good approximation in an experiment if, for all rest masses in the search range, the axion coherence length\cite{chaudhuri2018fundamental} is much larger than the extent of the magnetic field, which is nearly always the case\cite{sikivie2021invisible}. We assume that the currents driven by axion-induced fields occupy a finite volume. We address this assumption further at the end of Section \ref{ssec:impulse_correct}, in the context of shielding in experiments. 

We hold the axion dark-matter field to be a stiff source (with negligible source impedance), as justified in Section \ref{sec:gedank}. For a receiver that is well-described by classical axion electrodynamics (as opposed to quantum-mechanical descriptions, e.g., for Fock-state preparation in a cavity\cite{sikivie2021invisible,chou2019quantum}), we have\cite{Sik14,millar2017dielectric,kim2019effective,ouellet2019solutions,beutter2019axion}
\begin{equation}\label{eq:AED1}
    \vec{\nabla} \cdot \vec{E}= \rho/\epsilon_{0},\ \vec{\nabla} \times \vec{E}=-\partial_{t}\vec{B} 
\end{equation}
\begin{equation}\label{eq:AED2}
    \vec{\nabla} \cdot \vec{B}=0,\ \vec{\nabla} \times \vec{B} = \mu_{0} (\vec{J} + \vec{J}_{a}) + \mu_{0}\epsilon_{0} \partial_{t}\vec{E},
\end{equation}
where $\vec{E}$ and $\vec{B}$ are the electric and magnetic fields, respectively. The $\rho$ and $\vec{J}$ are the electromagnetic charge and current density. All fields and currents are functions of position $\vec{x}$ and time $t$. The arguments have been omitted for brevity. $\vec{J}_{a}$ is the effective axion current density, which models the fields produced by axion-to-photon conversion. It is related to the axion potential $a(t)$ by $\vec{J}_{a}(\vec{x},t)=- \kappa_{a}\vec{B}(\vec{x},t) \partial_{t}a(t)/Z_{fs}$.
 
Eqs. (\ref{eq:AED1}) and (\ref{eq:AED2}) contain information about the electromagnetic fields and currents arising from the axion field, the DC currents driving the DC magnetic field, noise fields and currents, as well as any other fields and currents in the receiver. To analyze impedance matching, we focus on the fields and currents arising from axions. Formally, one can perturbatively expand the fields and currents in orders of the axion-photon coupling $\kappa_{a}$:
\begin{equation}\label{eq:E_expand}
    \vec{E}(\vec{x},t)=\vec{E}_{0}(\vec{x},t) + \kappa_{a} \vec{E}_{1}(\vec{x},t) + \kappa_{a}^{2}\vec{E}_{2}(\vec{x},t) + ...
\end{equation}
and analogously for $\vec{B}(\vec{x},t)$, $\rho(\vec{x},t)$, and $\vec{J}(\vec{x},t)$. Terms with subscript '0' represent the fields and currents in the absence of the axion-photon interaction. We assume that the background DC magnetic field $\vec{B}_{DC}(\vec{x})$ is sufficiently large (e.g., much larger than thermal noise fields) such that $\vec{B}_{0}(\vec{x},t) \approx \vec{B}_{DC}(\vec{x})$. %We assume that $\vec{E}_{0}(\vec{x},t)$ is negligible. 
Since the axion field is stiff, we ignore terms that are order $\kappa_{a}^{2}$ and higher, representing the effects of backreaction. From here on, for brevity, we identify the fields and currents $\vec{E}(\vec{x},t)$, $\vec{B}(\vec{x},t)$, $\rho(\vec{x},t)$, $\vec{J}(\vec{x},t)$ with the first-order terms in the expansion. For example, $\vec{B}(\vec{x},t)$ is synonymous with $\kappa_{a}\vec{B}_{1}(\vec{x},t)$ and distinct from $\vec{B}_{DC}(\vec{x})$. Note that, in our thought experiment, $\vec{E}(\vec{x},t)$, now synonymous with $\kappa_{a}\vec{E}_{1}(\vec{x},t)$, represents the sum of the axion-induced background electric field $\vec{E}_{a}(t)$ and the fields produced by the receiver in response to this drive. We then have
\begin{equation}\label{eq:J_eff}
    \vec{J}_{a}(\vec{x},t) \approx -\frac{\kappa_{a}}{Z_{fs}} \vec{B}_{DC}(\vec{x}) \partial_{t}a(t).
\end{equation}

Consider a volume $V$, with boundary $\partial V$, surrounding the DC magnetic field and all currents driven by the axion-induced fields. From eqs. (\ref{eq:AED1}), (\ref{eq:AED2}), and (\ref{eq:J_eff}), 
\begin{equation}\label{eq:axion_Poynting}
    \frac{\kappa_{a}}{Z_{fs}}\partial_{t}a(t) \int_{V} \vec{E} \cdot \vec{B}_{DC} = P_{J}(t) + P_{rad}(t) + \partial_{t}(W_{B}+W_{E}),
\end{equation}
where
\begin{equation}\label{eq:UEM}
    W_{B}(t)= \frac{1}{2\mu_{0}} \int_{V} |\vec{B}(\vec{x},t)|^{2},\quad W_{E}(t)= \frac{\epsilon_{0}}{2} \int_{V}|\vec{E}(\vec{x},t)|^{2}
\end{equation}
\begin{equation}\label{eq:PJrad}
     P_{J}(t)=\int_{V} \vec{J}(\vec{x},t) \cdot \vec{E}(\vec{x},t),\quad P_{rad}(t)= \frac{1}{\mu_{0}} \int_{\partial V} (\vec{E}(\vec{x},t) \times \vec{B}(\vec{x},t)) \cdot \vec{da}.
\end{equation}
Eq. (\ref{eq:axion_Poynting}) is a statement of energy conservation. The rate of energy transfer from the axion field in volume $V$ is equal to the sum of: the rate of work done on electromagnetic charges, the electromagnetic energy flux through the surface $\partial V$, and the rate of change in field energy.\footnote{Note that in limiting $\vec{E}(\vec{x},t)$ to be the first-order field produced by the axion, we have ignored the possibility of coherent signal addition; if a monochromatic axion field sources an electric field that constructively interferes with a background electric field at the same frequency, the energy transfer from the axion is enhanced (relative to no such background electric field)\cite{chou2019quantum}. However, the phase of the axion dark-matter field, dictating the phase of the produced electric field, is unknown, so we ignore this scenario, which is not relevant to present laboratory searches.}

\subsection{Impedance Seen by the Axion Source}
\label{ssec:impedance}

%We now use (\ref{eq:axion_Poynting}) to formulate the concept of impedance seen by the axion field. 
Electromagnetic impedance relates an input current to a produced voltage\cite{fano1968electromagnetic}. We thus define the quantities that play the role of current and voltage in our analysis, after which we introduce assumptions underlying our concept of impedance and the gain-bandwidth bound:
\begin{equation}\label{eq:IVat}
    I_{a}(t) \equiv \frac{\kappa_{a}}{Z_{fs}}\partial_{t}a(t),\quad V_{a}(t) \equiv \int_{V} \vec{E}(\vec{x},t) \cdot \vec{B}_{DC}(\vec{x}).
\end{equation}

The quantity $I_{a}(t)$ plays the role of input current, being proportional to the effective current density (\ref{eq:J_eff}). %Its effect is governed by eqs. (\ref{eq:AED1}) and (\ref{eq:AED2}).
%\begin{equation}\label{eq:AED1_Iat}
%    \vec{\nabla} \cdot \vec{E}= \rho/\epsilon_{0},\ \vec{\nabla} \times \vec{E}=-\partial_{t}\vec{B} 
%\end{equation}
%\begin{equation}\label{eq:AED2_Iat}
%    \vec{\nabla} \cdot \vec{B}=0,\ \vec{\nabla} \times \vec{B} = \mu_{0} (\vec{J} -I_{a}(t) \vec{B}_{DC}(\vec{x})) + \mu_{0}\epsilon_{0} \partial_{t}\vec{E}.
%\end{equation}
%By studying the effect of $I_{a}(t)$ on producing electromagnetic fields and current flow, we can understand the fields and current flow induced by axions. In particular, 
To derive the gain-bandwidth bound, we will consider the effect of a delta function $I_{a}(t)$. A delta function is not a solution for the equation of motion of the axion field\cite{Sik83}. However, mathematically, the system response to a delta function reveals the behavior of axion-induced electromagnetic fields and currents over a wide range of possible rest-mass frequencies because its Fourier transform carries constant value in the frequency domain.

The quantity $V_{a}(t)$, characterizing the overlap of the electric field with the DC magnetic field, plays the role of voltage in our analysis. Note that the left-hand side of eq. (\ref{eq:axion_Poynting}) equals $I_{a}(t)V_{a}(t)$, similar to other source-power expressions in electromagnetic systems\cite{montgomery1987principles,fano1968electromagnetic}. We use the language of linear-response theory\cite{triverio2007stability} and refer to $I_{a}(t)$ as the input to the receiver and $V_{a}(t)$ as the output. 

We assume that the relationship between the input $I_{a}(t)$ and output $V_{a}(t)$ is linear and time-invariant. From eqs. (\ref{eq:AED1})-(\ref{eq:AED2}), one can observe that the assumption holds if the constitutive equations (e.g., susceptibility relations in a dielectric), describing the relationships between electromagnetic currents and fields in the system, are linear and time-invariant. 
We may then write
\begin{equation}\label{eq:Zdef}
    V_{a}(t)=\int_{-\infty}^{\infty}\ d\tau\ Z_{a}(t-\tau)I_{a}(\tau).
\end{equation}
We refer to the response function $Z_{a}(t)$ as the impedance seen by the axion field. The impedance describes the response of the photons and charges to an axion drive. 

We assume that, for all inputs $I_{a}(t)$ and all times $\tau$,
\begin{equation}\label{eq:EM_passive}
    \int_{-\infty}^{\tau} dt\ P_{J}(t) \geq 0.
\end{equation}
In other words, the charges do not supply energy in response to the axion-induced electromagnetic forcing. Eq. (\ref{eq:EM_passive}) holds for receivers containing only passive elements. Conversely, eq. (\ref{eq:EM_passive}) may not hold for receivers containing active elements\cite{clarke2006squid,caves1980measurement,chen2002circuits,wicht1997white,pati2007demonstration,salit2010enhancement,yum2013demonstration,nistad2008causality}. As per our description of active matching in Section \ref{sec:gedank}, the assumption (\ref{eq:EM_passive}) regarding charges is central to producing gain-bandwidth limitations.\footnote{At first glance, the assumption (\ref{eq:EM_passive}) seems to exclude laboratory receivers using amplifiers as readout elements. However, the amplifier's effect on the receiver impedance-match to axions is generally well-characterized by its input impedance, described as an equivalent-circuit passive load.\cite{castellanos2008amplification,hilbert1985measurements} We can apply our arguments to receivers with a readout amplifier by substituting this passive load.}

Because no electromagnetic sources exist beyond the volume $V$ by construction, the energy flow through $\partial V$ must be zero or net outwards:  for all inputs $I_{a}(t)$,
\begin{equation}\label{eq:rad_passive}
    \int_{-\infty}^{\tau} dt\ P_{\rm rad}(t) \geq 0.
\end{equation}
Combining (\ref{eq:axion_Poynting}), (\ref{eq:IVat}), (\ref{eq:EM_passive}), (\ref{eq:rad_passive}) with the fact that $W_{E}(t)$ and $W_{B}(t)$ are non-negative, we find that, for linear, time-invariant, passive receivers,  $\int_{-\infty}^{\tau} dt\ I_{a}(t)V_{a}(t) \geq 0$, i.e. energy is always extracted from axions.
%\begin{equation}\label{eq:ax_passive}
%    \int_{-\infty}^{\tau} dt\ I_{a}(t)V_{a}(t) \geq 0.
%\end{equation}
%That is, due to energy conservation and the assumption (\ref{eq:EM_passive}), energy is always extracted from the axions. 
$Z_{a}(t)$ therefore has a well-defined Fourier transform\cite{triverio2007stability}
\begin{equation}\label{eq:Z_FFT}
    Z_{a}(\omega)=\int_{-\infty}^{\infty} dt\ Z_{a}(t) \textrm{exp}(-i\omega t),
\end{equation}
and we may write eq. (\ref{eq:Zdef}) in the frequency domain as
\begin{equation}\label{eq:FFT_product}
    V_{a}(\omega)=Z_{a}(\omega)I_{a}(\omega).
\end{equation}

Denote the real part of the impedance $Z_{a}(\omega)$ as $R_{a}(\omega)$. Since the receiver is assumed to be passive, $R_{a}(\omega)$ is non-negative for all $\omega$\cite{triverio2007stability}. From Ch. 6.9 of ref. \cite{Jackson:1998nia}, we obtain
\begin{equation}\label{eq:Rdef}
    R_{a}(\omega) = \frac{1}{|I_{a}(\omega)|^{2}} \left( \int_{V} Re(\vec{J}^{*}(\vec{x},\omega) \cdot \vec{E}(\vec{x},\omega)) + \frac{1}{\mu_{0}}\int_{\partial V} Re(\vec{E}(\vec{x},\omega) \times \vec{B}^{*}(\vec{x},\omega)) \cdot \vec{da} \right)
\end{equation}
$\vec{E}(\vec{x},\omega)$ is the Fourier transform of $\vec{E}(\vec{x},t)$ and similarly for $\vec{J}$ and $\vec{B}$. $R_{a}(\omega)$ represents dissipation in charges and radiation due to an axion forcing at frequency $\omega$.

\subsection{Impulse Response}
\label{ssec:impulse_correct}

To constrain the frequency-integrated power absorption in a receiver, we must therefore constrain $R_{a}(\omega)$. We show that, for linear, time-invariant, passive receivers,
\begin{equation}\label{eq:BF_DM}
    \int_{0}^{\infty}d\omega\ R_{a}(\omega) \leq \frac{\pi}{2\epsilon_{0}} \int_{V} |\vec{B}_{DC}(\vec{x})|^{2}.
\end{equation}

We demonstrate (\ref{eq:BF_DM}) by placing an upper bound on the energy supplied to the system by an impulse\cite{lasenby2021parametrics}
\begin{equation}\label{eq:ax_impulse}
    I_{a}^{\delta}(t)=I_{a}^{0} \delta(t),\ I_{a}^{0}>0
\end{equation}
where $\delta(t)$ is the Dirac delta function. %The impulse has constant weight in frequency space---$I_{a}^{\delta}(\omega)=I_{a}^{0}$--and thus, provides information about the dissipative response over a wide range of possible rest-mass frequencies.
That an upper bound should exist follows from the intuition regarding the photon-electron interaction developed in Section \ref{sec:gedank}. The self-impedance of photons acting on charges limits current flow, which in turn, limits the electric field. A limitation on the electric field constrains the response $V_{a}^{\delta}(t)$. Since the energy supplied is given by the time integral of $I_{a}^{\delta}(t)V_{a}^{\delta}(t)$, we then expect a limit on its value. 
%The limitation leads to eq. (\ref{eq:BF_DM}). %Ref. \cite{lasenby2021parametrics}, which introduces impulse response, claims a gain-bandwidth relation similar to (\ref{eq:BF_DM}), but the argument is incomplete because it does not apply to systems with electromagnetic charges. See Appendix \ref{ssec:imp_response} for details.

It is important to understand the limits of integration $(0,+\infty)$ in eq. (\ref{eq:BF_DM}), which is a product of mathematical extrapolation rather than physical deduction. We are not making assumptions about the photon-electron interaction at arbitrarily high or low frequency, but rather working from the understanding of the interaction at the frequency scales relevant to searches (e.g., $\lesssim$200 MHz for lumped-element searches, $\gtrsim$600 MHz for cavity searches\cite{sikivie2021invisible}). At these scales, the observed relationships between electromagnetic fields and currents in a linear, time-invariant, passive receiver are well-described by linear, time-invariant constitutive equations, e.g., Ohm's Law and descriptions of dielectrics using susceptibility tensors. These equations are compatible with the usage of linear response in eq. (\ref{eq:Zdef}) and are consistent with eq. (\ref{eq:EM_passive}). Plugging the constitutive equations into (\ref{eq:axion_Poynting}), we mathematically infer a bound on the energy transfer from the impulse (\ref{eq:ax_impulse}), which contains all frequencies. This leads to (\ref{eq:BF_DM}). We then deduce a maximum on axion signal power, integrated over all possible rest-mass frequencies in the search range, leading to (\ref{eq:PJ_DM}) below. In other words, though we mathematically extrapolate the constitutive relations to outside the search range (where they fail physically) in order to establish (\ref{eq:BF_DM}), we can still use eq. (\ref{eq:BF_DM}) to constrain power flow in the search range since the relations are physically valid there. For what follows, we denote all response quantities resulting from the pulse with a superscript $\delta$.\footnote{Our approach is similar to that used in establishing the Bode-Fano criterion\cite{bode1945network,fano1950theoretical}. Consider a two-port system that, over a frequency range of interest, can be described by a lumped-element equivalent circuit with a source resistor and a load consisting of a series inductor and resistor. Under certain assumptions, the Bode-Fano criterion constrains the gain-bandwidth product of the match between the source and the load. In proving the criterion, one uses the voltage-current relationships governing lumped elements to determine the set of possible impedance functions that can be seen by the load resistor. Based on this set, one mathematically infers the behavior of the impedance function in the limit of infinite frequency (which is essentially dual to inferring behavior on arbitrarily short time scales, as in an impulse-response argument). The result is used to mathematically constrain the pole structure of the circuit's return-loss function in the complex plane. One then deduces the maximum gain-bandwidth product of the two-port system in the frequency range over which the circuit is an accurate physical descriptor.}

Consider a time $\tau>0$. Integrating eq. (\ref{eq:axion_Poynting}) yields, for the energy supplied by the pulse over all time,
\begin{align}
    W_{a}^{\delta}& =\int_{-\infty}^{\infty} dt\ I_{a}^{\delta}(t)V_{a}^{\delta}(t)=\int_{-\infty}^{\tau} dt\ I_{a}^{\delta}(t)V_{a}^{\delta}(t) \label{eq:work_LB} \\
    &=  \int_{-\infty}^{\tau} dt\ [P_{J}^{\delta}(t) + P_{rad}^{\delta}(t) + \partial_{t}(W_{B}^{\delta}+W_{E}^{\delta})] \nonumber \\
    & \geq \frac{\epsilon_{0}}{2} \int_{V} |\vec{E}^{\delta}(\vec{x},\tau)|^{2} \geq \frac{\epsilon_{0}}{2} \frac{(\int_{V} \vec{E}^{\delta}(\vec{x},\tau) \cdot \vec{B}_{DC}(\vec{x}))^{2}}{\int_{V}|\vec{B}_{DC}(\vec{x})|^{2} }. \nonumber
\end{align}
In the first inequality, we have used eqs. (\ref{eq:EM_passive})-(\ref{eq:rad_passive}). 
%We have also used the fact that $W_{B}$ is non-negative at time $\tau$ and that the electric and magnetic fields vanish in the limit $t \rightarrow -\infty$. 
We have used the Cauchy-Schwarz inequality in the last step.

For the pulse (\ref{eq:ax_impulse}), eq. (\ref{eq:IVat}) yields
\begin{equation}
    W_{a}^{\delta} = I_{a}^{0} \int_{V} \frac{\vec{E}^{\delta}(\vec{x},0+) + \vec{E}^{\delta}(\vec{x},0-)}{2} \cdot \vec{B}_{DC}(\vec{x}). \label{eq:work_axion}
\end{equation}
$\vec{E}^{\delta}(\vec{x},0+) \equiv \lim_{t\to 0 +} \vec{E}^{\delta}(\vec{x},t)$ is the electric field immediately after the pulse, and $\vec{E}^{\delta}(\vec{x},0-)$ is the field immediately before the pulse. The latter vanishes. Plugging (\ref{eq:work_axion}) into (\ref{eq:work_LB}) and taking the limit $\tau \rightarrow 0+$ then implies
\begin{equation}\label{eq:work_UB}
    W_{a}^{\delta} \leq \frac{(I_{a}^{0})^{2}}{2\epsilon_{0}} \int_{V} |\vec{B}_{DC}(\vec{x})|^{2}.
\end{equation}
Fourier-transforming the first integrand of (\ref{eq:work_LB}), with $I_{a}^{\delta}(\omega)=I_{a}^{0}$ for all $\omega$, gives, using eq. (\ref{eq:FFT_product}),
\begin{equation}
    W_{a}^{\delta} =\frac{(I_{a}^{0})^{2}}{2\pi}\int_{-\infty}^{\infty} d\omega\ Z_{a}(\omega) = \frac{(I_{a}^{0})^{2}}{\pi} \int_{0}^{\infty} d\omega\ R_{a}(\omega) \label{eq:work_GBP}
\end{equation}
We have used that $Z_{a}(\omega)=Z_{a}^{*}(-\omega)$ since $Z_{a}(t)$ is real-valued. Combining (\ref{eq:work_UB}) and (\ref{eq:work_GBP}) yields the desired (\ref{eq:BF_DM}).

To show explicitly how (\ref{eq:BF_DM}) constrains power absorption, we consider a situation resembling a laboratory search between possible rest-mass frequencies $\omega_{l}$ and $\omega_{h}$. We calculate an upper bound on axion-field power dissipated in the receiver, integrated over this search band. We assume that for all frequencies in the search, the coherence length is much larger than the extent of the DC magnetic field, so that the axion field is spatially uniform and (\ref{eq:BF_DM}) applies. We hold fixed the axion-photon coupling, dark-matter energy density $\rho_{a}$, and DC magnetic field, over the search range. For simplicity, we assume a monochromatic, rather than virialized, axion field, as expressed in (\ref{eq:axion_field}). We stress however that (\ref{eq:BF_DM}) constrains power flow generally, regardless of the specific assumptions about the cold-dark-matter velocity distribution.

Following eq. (\ref{eq:IVat}), the instantaneous power flow is $I_{a}(t)V_{a}(t)$, so eqs. (\ref{eq:rho_a}), (\ref{eq:Zdef}), (\ref{eq:FFT_product}), and (\ref{eq:Rdef}) give
\begin{equation}\label{eq:Pdiss_FFT}
    P(\omega_{a}) \equiv \frac{(\kappa_{a}c)^{2}}{\mu_{0}} \rho_{a} R_{a}(\omega_{a})
\end{equation}
as the time-averaged power flow from an axion field of rest-mass frequency $\omega_{a}$. Integrating using eq. (\ref{eq:BF_DM}), 
\begin{equation}
    \int_{\omega_{l}}^{\omega_{h}} d\omega_{a}\ P(\omega_{a}) \leq \frac{\pi}{2} (\kappa_{a}c^{2})^{2} \rho_{a} \int_{V} |\vec{B}_{DC}(\vec{x})|^{2}.\label{eq:intP_DM}
\end{equation} 
Denoting $P_{J}(\omega_{a})$ as the power dissipated in the charges, we have $P_{J}(\omega_{a}) \leq P(\omega_{a})$, so
\begin{equation}
    \int_{\omega_{l}}^{\omega_{h}} d\omega_{a}\ P_{J}(\omega_{a}) \leq \frac{\pi}{2} (\kappa_{a}c^{2})^{2} \rho_{a} \int_{V} |\vec{B}_{DC}(\vec{x})|^{2}.\label{eq:PJ_DM}
\end{equation} 

The power dissipated in the receiver, integrated over all rest-mass frequencies, is proportional to $\kappa_{a}^{2}$, but, as discussed, the power available from dark matter is independent of $\kappa_{a}$. Thus, for practical background magnetic-field strengths, owing to limitations from the photon-electron interaction, an efficient, broadband impedance match to a range of possible rest-mass frequencies in a linear, time-invariant, passive receiver is fundamentally impossible. We will further consider the practical consequences of the photon-electron interaction and the resulting bound on gain-bandwidth product in Section \ref{sec:implications}.
%If a receiver is designed to be an efficient match at one possible rest-mass frequency, the match at other possible frequencies is necessarily limited.

%\sch{Is the nuance presented in this paragraph too far in the weeds?} 
As a final remark for this section and as discussed at the beginning of Section \ref{ssec:COE}, our derivation of (\ref{eq:intP_DM}) and (\ref{eq:PJ_DM}) relies on the axion-driven currents occupying a finite volume. This condition may be difficult to guarantee in an experiment, especially if axion-induced photons radiate into free space and interact with charges far from the apparatus. However, the experimentalist is usually only interested in measuring the signal power deposited in some sub-volume of charges, e.g., the power delivered to the input of an amplifier, rather than the power also dissipated in magnet wires and mechanical supports. The sub-volume is often well-shielded to mitigate electromagnetic interference, which can degrade sensitivity. Suppose that the sub-volume of interest is contained within a volume $V_{s}$. We show how the bounds (\ref{eq:intP_DM})-(\ref{eq:PJ_DM}) can still apply to the volume $V_{s}$. 

We modify the definition of the output $V_{a}(t)$ in (\ref{eq:IVat}) to be an integral over $V_{s}$. We assume that the relationship between $I_{a}(t)$ and $V_{a}(t)$ is linear and time-invariant and that, within the volume $V_{s}$, the receiver possesses only passive elements so that (\ref{eq:EM_passive}) is satisfied. Suppose also that eq. (\ref{eq:rad_passive}) is satisfied over the boundary $\partial V_{s}$, as justified below. Then, using the impulse-response approach, we find integrated-power bounds similar to (\ref{eq:intP_DM}) and (\ref{eq:PJ_DM}), except that the volume integrals are performed over $V_{s}$. Since $V_{s}$ may or may not entirely contain the DC magnetic field, eqs. (\ref{eq:intP_DM}) and (\ref{eq:PJ_DM}) are also satisfied. 

Eq. (\ref{eq:rad_passive}) is satisfied in general laboratory situations. For instance, suppose the receiver is completely shielded by a high-conductivity normal (superconducting) metal, several skin (penetration) depths thick at all frequencies within the search band. Taking the volume boundary $\partial V_{s}$ to be within the bulk of the shield, the electric field on the boundary vanishes and (\ref{eq:rad_passive}) is satisfied with equality.\footnote{Note that, in studying the impulse response, we extrapolate the boundary conditions, describing the electric field within the shield bulk as vanishingly small, to frequencies beyond the search range. The extrapolation may fail physically, e.g., at low frequencies for normal-metal shields, where the skin depth can be arbitrarily large. However, as discussed earlier, we may still make deductions applying to the search band.} Similar considerations, showing (\ref{eq:rad_passive}) is satsified, may be applied to a cavity resonator connected to a $Z_{0}=$50 $\Omega$ readout waveguide, for which the power flow is $P_{rad}(t)=I(t)^{2}Z_{0}\geq 0$ where $I(t)$ is the waveguide current.

%------------------------------------------------------

\section{Optimizing the Impedance Match}
\label{sec:search_opt}

We determine receivers that can saturate the bound (\ref{eq:PJ_DM}) on the match to the axion dark matter source, imposed by the photon-electron coupling. The results of this section are central to the discussion of practical implications of our analysis in Section \ref{sec:implications}. 

We first demonstrate that a single-pole, high-$Q$ resonant mode of a cavity receiver, widely used in axion dark-matter searches, can saturate the bound. Consider a free-space cavity of volume $V_{c}$ with negligible participation from dielectric or magnetic materials. Suppose the cavity possesses a high-$Q$ resonant mode at center frequency $\omega_{c}$ within the given search band, $\omega_{l} < \omega_{c} < \omega_{h}$. Suppose that the mode is characterized by a complex-valued electric-field mode function $\vec{E}_{c}(\vec{x})$ and that the search bandwidth is at least several resonator bandwidths on each side of the center frequency. Within the search range, the axion field is assumed to be approximately monochromatic and spatially uniform over the DC magnetic field extent. 
Assume that the mode is well-separated from other cavity modes (so that we may ignore mode interactions). Then, for possible axion rest-mass frequencies $\omega_{a}$ satisfying $|\omega_{a}-\omega_{c}| \ll \omega_{c}$,\cite{Sik85,krauss1985calculations,Graham:2014sha,peng2000cryogenic}
\begin{equation}
    P_{J,c}(\omega_{a}) \approx (\kappa_{a}c^{2})^{2} \rho_{a}  \mathcal{C}  \frac{Q}{\omega_{c}} \frac{\int_{V_{c}} |\vec{B}_{DC}(\vec{x})|^{2}}{1+4Q^{2}(\omega_{a}-\omega_{c})^{2}/\omega_{c}^{2}},\label{eq:Pjc}
\end{equation}
is the power dissipation in the cavity, where
\begin{equation}
    \mathcal{C}= \frac{|\int_{V_{c}} \vec{E}_{c}^{*}(\vec{x}) \cdot \vec{B}_{DC}(\vec{x}) |^{2}} {\int_{V_{c}} |\vec{E}_{c}(\vec{x})|^{2} \int_{V_{c}} |\vec{B}_{DC}(\vec{x})|^{2}}\label{eq:cav_overlap}
\end{equation}
describes the mode function overlap with the DC field, with a maximum possible value of unity (although in practice it is less than unity \cite{du2018search}). 
Integrating, we find
\begin{equation}
    \int_{\omega_{l}}^{\omega_{h}} d\omega_{a}\ P_{J,c}(\omega_{a}) \approx \frac{\pi}{2} (\kappa_{a}c^{2})^{2} \rho_{a} \mathcal{C} \int_{V_{c}} |\vec{B}_{DC}(\vec{x})|^{2}. \label{eq:Pjc_int}
\end{equation}
In the limit that the cavity encloses the entire DC magnetic field and $\mathcal{C}$ approaches unity, we recover the upper bound of (\ref{eq:PJ_DM}). %The high-Q, single-pole cavity may therefore, in theory, saturate the gain-bandwidth bound of eq. (\ref{eq:PJ_DM}).
Note that the bound may also be saturated when summed over multiple cavity modes, instead of a single mode with unity overlap factor. This is the case with a high-$Q$ cylindrical cavity containing uniform DC magnetic field along its central axis. See eq. (70) of ref. \cite{sikivie2021invisible} and eqs. (5.7) and (5.18) of ref. \cite{steiner1987spectral}. 

However, note that the frequency-integrated power dissipation (\ref{eq:Pjc_int}) is independent of $Q$. Moreover, in Section \ref{sec:gedank}, one may observe that the integrated power dissipation of the cavity mode and the broadband antenna, in a range of several bandwidths about the resonance, are comparable. This suggests that there may be may be large classes of receivers that saturate the gain-bandwidth bound (\ref{eq:PJ_DM}) on the impedance match to dark matter.

\subsection{Optimizing with the Bode Integral Theorem}
\label{ssec:Bode}

%A systematic framework for determining receivers that saturate (\ref{eq:PJ_DM}) is obtained through the use of equivalent circuits. 
We investigate single-moded receivers that are inductively coupled to the axion signal, i.e. which, over the given search range bounded by $\omega_{l}$ and $\omega_{h}$, can be modeled as driven through a single equivalent-circuit inductance. This model describes, among other things, a single mode in a typical cavity haloscope\cite{montgomery1987principles,khatiwada2020axion}. (Nevertheless, we stress that the gain-bandwidth bound also applies for multi-moded receivers, as discussed further in Section \ref{sec:implications}.) Using the gain-bandwidth bounds derived in Section \ref{sec:GBP}, we show that there is a maximum amplitude for the equivalent-circuit voltage source parametrizing the receiver excitation. We then combine the voltage constraint with the Bode integral theorem\cite{bode1945network,fano1950theoretical}, which constrains equivalent-circuit impedance and is independent of the derivations thus far, to bound power dissipation integrated over the search range. The bound is equivalent to eq. (\ref{eq:PJ_DM}); the equivalence yields conditions for receivers to saturate the gain-bandwidth bound on power dissipation over a given search range and classes of receivers that can meet these conditions.

\begin{figure}[htp] 
\centering
\includegraphics[width=0.6\textwidth]{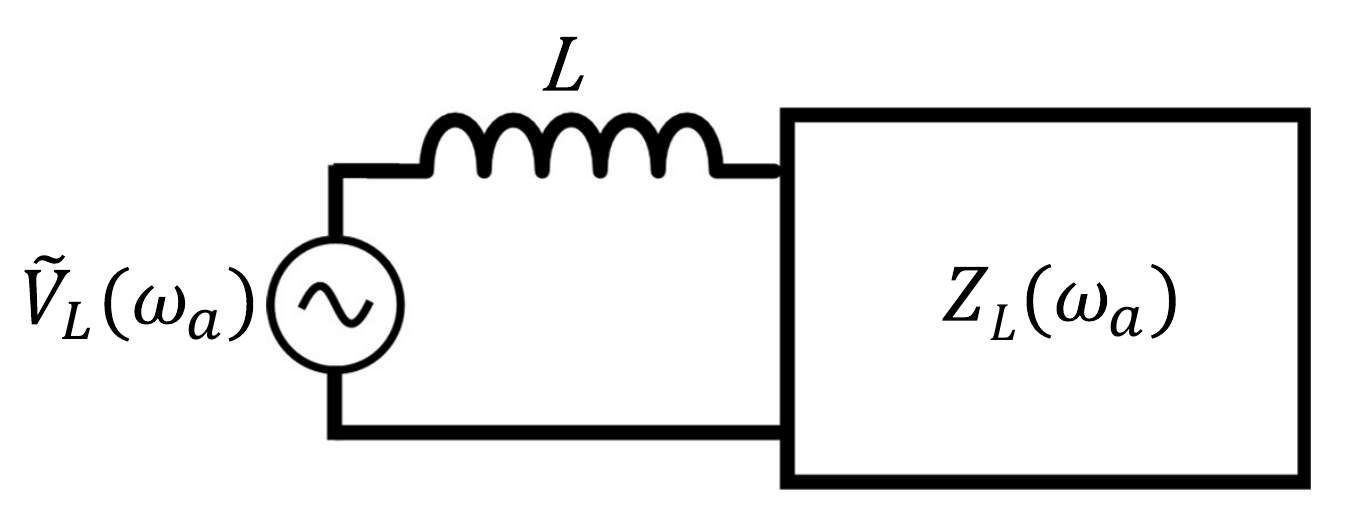}
\caption{Equivalent circuit of a receiver coupled to the axion excitation $\tilde{V}_{L}$ through an equivalent inductance $L$.  \label{fig:IndBF}}
\end{figure}
%The axion drive of rest-mass frequency $\omega_{a}$ is represented by a voltage phasor $\tilde{V}_{L}(\omega_{a})$.

See Fig. \ref{fig:IndBF}. 
%The axion field of rest-mass frequency $\omega_{a}$ produces electromagnetic fields driving a stiff voltage $V_{L}(t)$ across an inductance $L$. 
The equivalent-circuit inductance is attached to a linear, passive load impedance $Z_{L}$, which we assume to consist of inductors, capacitors, and/or resistors with frequency-independent values. So that we may compare the integrated-power bound derived from the equivalent-circuit approach to eq. (\ref{eq:PJ_DM}), we assume that the resistors represent solely dissipation in charges, rather than radiation loss. Additionally, receiver resistances physically tend to have frequency-dependent value, but over a finite search bandwidth or especially within a narrow bandwidth $\omega_{h}-\omega_{l} \ll \omega_{l}$, can often be approximated as frequency-independent, which we assume here. As before, we consider a monochromatic axion field (\ref{eq:axion_field}), resulting in a voltage drive characterized by complex phasor $\tilde{V}_{L}(\omega_{a})$. In the equivalent circuit, we ignore the axion source impedance, demonstrated to be negligible in Section \ref{sec:gedank}. The power delivered to the load is
\begin{equation}\label{eq:PL_bound}
    P_{L}(\omega_{a}) =\frac{1}{2} \frac{|\tilde{V}_{L}(\omega_{a})|^{2}}{|Z_{T}(\omega_{a})|^{2}} Re(Z_{T}(\omega_{a})),
\end{equation}
where $Z_{T}(\omega_{a})=i\omega_{a}L +Z_{L}(\omega_{a})$ is the circuit impedance.
%\begin{equation}\label{eq:ZT_def}
%    Z_{T}(\omega_{a})=i\omega_{a}L +Z_{L}(\omega_{a})
%\end{equation}
%seen by the axion-induced voltage source at frequency $\omega_{a}$. 

%As Schwinger states, Maxwell's equations are used to constrain the parameter values in equivalent circuits\cite{schwinger1969julian,milton2007appreciation}. Thus, 
We first note that, for any rest-mass frequency $\omega_{a}$ within the search range, the voltage phasor obeys
\begin{equation}\label{eq:VL_ineq}
    |\tilde{V}_{L}(\omega_{a})|^{2} \leq 2L (\kappa_{a}c^{2})^{2} \rho_{a} \int_{V} |\vec{B}_{DC}(\vec{x})|^{2}.
\end{equation}
where $V$ surrounds the background DC magnetic field.\footnote{This implies $k(\omega_{\rm DM}^{0})^{2} \leq 1/2$ in Section III B of ref. \cite{chaudhuri2018fundamental}. Note that treatments similar to that in Section \ref{sec:search_opt}A can be performed for capacitively coupled receivers.} To see this, assume that the square modulus of the voltage phasor exceeds the bound at some frequency $\omega_{a}^{*}$ and that $\tilde{V}_{L}$ is a continuous function of rest-mass frequency (a reasonable assumption for a physical system). %\sch{I think this is a reasonable assumption for a physical system.} 
Then, one can construct a high-$Q$, equivalent cavity circuit, with series resistor and capacitor in the load $Z_{L}$ and with resonance frequency $\omega_{a}^{*}$, such that the frequency-integrated power delivered to the load exceeds the bound (\ref{eq:PJ_DM}). Such is not possible for a linear, time-invariant, passive receiver, so eq. (\ref{eq:VL_ineq}) must hold.

We demonstrate that
\begin{equation}\label{eq:Bode_P}
    \int_{0}^{\infty} d\omega_{a} \frac{L Re(Z_{T}(\omega_{a}))}{ |Z_{T}(\omega_{a})|^{2} } \leq \frac{\pi}{2}.
\end{equation}
We extend $Z_{T}(\omega)$ to the complex plane:\cite{fano1968electromagnetic}
\begin{equation}
    Z_{T}(\omega) \rightarrow \tilde{Z}_{T}(p),\quad \tilde{Z}_{T}(p=i\omega)=Z_{T}(\omega).
\end{equation}
By construction, $\tilde{Z}_{T}(p)$ is a real-rational function. Consider an arbitrary resistance $R_{s}>0$ and the reflectance
\begin{equation}\label{eq:Gp_def}
    \Gamma(p,R_{s})=\frac{\tilde{Z}_{T}(p)-R_{s}}{\tilde{Z}_{T}(p)+R_{s}}.
\end{equation}
We also define $\zeta(\omega,R_{s}) \equiv 1- |\Gamma(p=i\omega,R_{s})|^{2}$. The Bode integral theorem gives\cite{bode1945network,fano1950theoretical}
\begin{equation}\label{eq:Bode_thm}
    \int_{0}^{\infty} d\omega_{a}\ \textrm{ln} \left( \frac{1}{|\Gamma(i\omega_{a},R_{s})|} \right) = \frac{\pi R_{s}}{L} - \pi \sum_{m} p_{m}(R_{s})
\end{equation}
where $\{p_{m}(R_{s})\}$ represent the zeros of $\Gamma(p,R_{s})$ in the right-half of the complex plane. Because $\tilde{Z}_{T}(p)$ is real-rational, the zeros must occur in pairs which are complex conjugates, so the sum in (\ref{eq:Bode_thm}) is non-negative. Thus,
\begin{equation}\label{eq:Bode_zeta}
    \int_{0}^{\infty} d\omega_{a}\ \frac{L}{4R_{s}}\textrm{ln} \left( \frac{1}{1-\zeta(\omega_{a},R_{s})} \right) \leq \frac{\pi}{2}.
\end{equation}
Since $0 \leq \zeta(\omega_{a},R_{s}) \leq 1 $, $\zeta \leq \textrm{ln}((1-\zeta)^{-1})$ and
\begin{equation}\label{eq:zeta_int}
    \int_{0}^{\infty} d\omega_{a}\ \frac{L Re(Z_{T}(\omega_{a}))}{ |Z_{T}(\omega_{a}) +R_{s}|^{2} }= \int_{0}^{\infty} d\omega_{a}\ \frac{L}{4R_{s}}\zeta(\omega_{a},R_{s}) \leq \frac{ \pi}{2}.
\end{equation}
Eq. (\ref{eq:zeta_int}) holds for arbitrary $R_{s}>0$, so (\ref{eq:Bode_P}) must hold. 

From eqs. (\ref{eq:PL_bound})-(\ref{eq:Bode_P}), the power delivered to the load, integrated over the search range $\omega_{l} \leq \omega_{a} \leq \omega_{h}$, satisfies
\begin{align}
    \int_{\omega_{l}}^{\omega_{h}} d\omega_{a}\ P_{L}(\omega_{a}) & \leq (\kappa_{a}c^{2})^{2} \rho_{a} \int_{V} |\vec{B}_{DC}(\vec{x})|^{2} \int_{\omega_{l}}^{\omega_{h}} d\omega_{a} \frac{L Re(Z_{T}(\omega_{a}))}{ |Z_{T}(\omega_{a})|^{2} } \nonumber \\
    &\leq (\kappa_{a}c^{2})^{2} \rho_{a} \int_{V} |\vec{B}_{DC}(\vec{x})|^{2} \int_{0}^{\infty} d\omega_{a} \frac{L Re(Z_{T}(\omega_{a}))}{ |Z_{T}(\omega_{a})|^{2} } \nonumber \\
    &\leq \frac{\pi}{2}(\kappa_{a}c^{2})^{2} \rho_{a} \int_{V} |\vec{B}_{DC}(\vec{x})|^{2},\label{eq:intPL_bound}
\end{align}
which is equivalent to the bound of (\ref{eq:PJ_DM}). 

The series of inequalities demonstrates that a receiver, consisting of equivalent resistors, inductors, and/or capacitors with frequency-independent value, may saturate the integrated-power bound over the search range if:
\begin{enumerate}
    \item For all frequencies at which the power delivered to the load $Z_{L}(\omega_{a})$ is non-negligible, the voltage constraint (\ref{eq:VL_ineq}) is saturated (first inequality).
    \item Outside of the search range, the frequency-integrated power delivered to the load $Z_{L}$ is negligible (second inequality).
    \item The inequality (\ref{eq:Bode_P}), derived from the Bode theorem, is saturated (third inequality).
\end{enumerate}

Condition 1 cannot be satisfied in the quasi-static limit, in which the spatial extent of the receiver and DC magnetic field is much smaller than the Compton wavelength. This is the regime of low-mass searches DM Radio and ABRACADABRA\cite{Chaudhuri:2014dla,kahn2016broadband,Silva-Feaver:2016qhh,ouellet2019first,phipps2020exclusion,salemi2021search}. However, as shown in \cite{chaudhuri2018fundamental}, the condition may be satisfied when the size of the receiver and DC magnetic-field extent is comparable to or larger than the Compton wavelength, e.g., as in ADMX and HAYSTAC cavity haloscopes\cite{brubaker2017first,du2018search,braine2020extended,backes2021quantum}. For a free-space cavity mode entirely containing the DC magnetic field, the ratio of $|\tilde{V}_{L}(\omega_{a})|^{2}$ to its maximum (\ref{eq:VL_ineq}) is the overlap factor (\ref{eq:cav_overlap}), which can, in theory, approach unity. 
Condition 2 can be satisfied by engineering the bandwidth of the receiver frequency response (e.g., by constructing a sufficiently narrowband receiver).

To understand how condition 3 may be satisfied, suppose that the load $Z_{L}(\omega_{a})$ possesses as its first element a frequency-independent series resistance $R$, so that
\begin{equation}\label{eq:ZT_prime}
    \tilde{Z}_{T}(p)= pL + R + \tilde{Z}'_{T}(p),
\end{equation}
where $\tilde{Z}'_{T}(p)$ is another real-rational, passive impedance.
The real part of $\tilde{Z}'_{T}(p)+pL$ must then have non-negative value on the right-half of the complex plane\cite{fano1968electromagnetic}. Then, for $R_{s} < R$, the reflectance (\ref{eq:Gp_def}) possesses no zeros in that domain and the sum in (\ref{eq:Bode_thm}) vanishes. Since
\begin{equation}
    \lim_{R_{s} \to 0+} \left[ \frac{L}{4R_{s}}\textrm{ln} \left( \frac{1}{1-\zeta(\omega_{a},R_{s})} \right) - \frac{L}{4R_{s}} \zeta(\omega_{a},R_{s}) \right] =0
\end{equation}
 for each frequency $\omega_{a}$, we have from eq. (\ref{eq:Bode_thm}),
\begin{equation}
    \int_{0}^{\infty} d\omega_{a} \frac{L Re(Z_{T}(\omega_{a}))}{ |Z_{T}(\omega_{a})|^{2} } =\int_{0}^{\infty} d\omega_{a}\ \lim_{R_{s} \to 0+}  \frac{L}{4R_{s}} \zeta(\omega_{a},R_{s}) = \frac{\pi}{2}.\label{eq:Bode_zeta_limRs}
\end{equation}
Condition 3 is thus satisfied if the equivalent-circuit inductive pickup is accompanied by a frequency-independent series resistance, often a good approximation in a narrowband receiver. If the series resistance representing loss possesses significant frequency dependence over the search range (e.g., as might be the case for a broadband receiver at high frequencies) or if the load impedance possesses active elements, then the statements here need not apply. 

Thus, we find that the gain-bandwidth bound (\ref{eq:PJ_DM}) can be saturated by a single-pole resonant mode of a cavity, with overlap factor approaching unity, as well as by, e.g., narrowband Bessel and Chebyshev waveguide filters\cite{matthaei1980microwave}.

%------------------------------------------------------

\section{Implications for Experiments}
\label{sec:implications}

For fixed DC magnetic field, subject to the constraint that the receiver be linear, time-invariant, and passive, the photon-electron interaction and consequent gain-bandwidth bound dictate that one cannot improve upon the optimized cavity haloscope in frequency-integrated power absorption. This has a number of interesting fundamental consequences. For example, it is impossible to produce a linear, time-invariant, passive axion receiver that simultaneously resonates (zero self-impedance) at all rest-mass frequencies. Additionally, one cannot evade the gain-bandwidth bound by subdividing the DC magnetic field volume into multiple linear, time-invariant, passive receivers and passively combining their outputs. To improve absorption beyond the gain-bandwidth bound, one must relax the underlying assumptions. One possibility is active receivers\cite{aberle2008two,rybka2014improving,daw2019resonant}--in which electrons supply energy--although they are more complex than the passive systems discussed here.

%This is usually accompanied by a large increase in implementation complexity, e.g., resulting from potential instabilities in active receivers\cite{aberle2008two}.

At the same time, our analysis can be used to guide practical improvements in linear, time-invariant, passive axion searches. It is well-known that difficulties emerge in constructing high-overlap-factor cavity haloscopes of volume much larger than the cube of the axion Compton wavelength. 
This has motivated the design of multi-pole coupled cavity filters that do not suffer from mode-mixing\cite{melcon2020scalable}. Our analysis in Section \ref{ssec:Bode} yields sufficient conditions for multi-pole filters to saturate the bound on integrated signal power. At low masses $\lesssim$1 $\mu$eV, a free-space cavity comparable in size to the Compton wavelength is too large to be practical, so lumped circuits are used as axion receivers\cite{sikivie2021invisible}. The constraint on voltage (\ref{eq:VL_ineq}) can be used to check calculations of receiver excitation in these circuits, which is particularly useful at sufficiently high rest-mass frequencies where signal dephasing and deviations from the quasi-static limit can introduce complexity (e.g., partially the regime of DMRadio-m$^{3}$\cite{ouelletsnowmass2021}).

More broadly, the practical difficulties with traditional cavity haloscopes at high masses and at low masses have motivated a large number of new techniques, including lumped-LC resonators, cavity arrays with power combiners, dielectric haloscopes, large-area antennas, and tunable plasmas\cite{Chaudhuri:2014dla,kahn2016broadband,horns2013searching,jeong2018phase,caldwell2017dielectric,lawson2019tunable}. The derived gain-bandwidth bound (\ref{eq:PJ_DM}) can be used to check receiver calibrations and calculations of power absorption, which is useful for gaining an understanding of achievable signal in any new architecture. The bound is especially valuable for multi-wavelength and multi-moded receivers, for which analytical approaches tend to be limited and numerical techniques are often computationally intensive.

There are significant apparent tensions between the bound (\ref{eq:PJ_DM}) and the projected power absorption in several proposed experiments. Ref. \cite{tobar2020broadband} claims a design evading the quasi-static suppression (see Section \ref{ssec:Bode} as well as refs. \cite{chaudhuri2018fundamental,ouellet2019first}), with the absorbed power diverging as the rest-mass frequency approaches zero. See eq. (35) of that work. However, the gain-bandwidth bound seemingly prohibits such a divergence. Refs. \cite{horns2013searching,irastorza2018new,BREAD2021,BREAD2021v2} discuss axion searches with a broadband dish antenna, in which the signal power scales linearly with the area of the dish, similar to eq. (\ref{eq:Pdiss_sheet}). It is thus claimed that a large area can produce significant sensitivity over a many-octave range of rest-mass frequencies without tuning. See, e.g., eqs. (2.19) and (3.3) of \cite{horns2013searching} or eqs (7.14)-(7.15) of \cite{irastorza2018new}. However, the limit on frequency-integrated power absorption scales as the volume integral of DC magnetic-field energy density; the difference in dimensionality of area and volume seems to imply that the projections in \cite{horns2013searching,irastorza2018new,BREAD2021,BREAD2021v2} fail at some rest-mass frequencies. (For example, consider a flat dish of area $A_{dish}$ immersed in a uniform magnetic field, tangent to its surface and of volume $V_{mag}$. Then, from eqs. (2.19) and (3.3) of \cite{horns2013searching} and eq. (\ref{eq:intP_DM}) of our work, the gain-bandwidth bound is violated if $2A_{dish}/\pi > V_{mag}(c/\omega_{l} - c/\omega_{h})^{-1}$. The violation thus occurs for appropriate sets of search frequency limits.) At sufficiently low frequency, this apparent tension arises from quasi-static physics, which adds suppression factors to the signal power computed in \cite{horns2013searching,irastorza2018new,BREAD2021,BREAD2021v2}. At higher frequency, when the linear dimension of the dish is multiple Compton wavelengths, one must account for coupling of the dish to cavity modes of the surrounding shield and interactions with fields generated due to the magnet boundary\cite{millar2017dielectric}. These inherently add frequency-dependent geometric factors to the signal power, increasing sensitivity at some frequencies, decreasing it at others, and thus changing the broadband nature of performance. One thus observes that the gain-bandwidth bound can be used to improve the accuracy of previous dish antenna analyses because it applies rigorously in the presence of complex interference effects caused by the environment. A full, quantitative resolution of the apparent tension is left to future work, as it requires a more detailed specification of magnet and receiver geometries.

%These considerations may also be relevant for similar multi-wavelength experiments claiming an absorption scaling with area, such as dielectric haloscopes\cite{caldwell2017dielectric,baryakhtar2018axion}. Indeed, the MADMAX effort has identified diffraction and interference effects as an important topic for investigation\cite{egge2020first,knirck2021simulating}. 

%One must also consider interactions between return loss at the focus, where the amplifier/photon counter is located, and the reflector, which result in additional frequency-dependent geometric factors. 

Furthermore, because the self-impedance of the photon-electron coupling governs absorption from axions, one should consider all objects (as all objects contain electrons) in the experimental volume for most accurately calculating signal power in a practical open system that is many Compton wavelengths in size. 
%it is often not sufficient to consider only the target object (such as the dish antenna) that is intended to enhance the axion signal. Because the self-impedance of the photon-electron coupling governs absorption from axions, one should instead account for all objects in the experimental volume, as they all contain electrons. 
Our analysis thus motivates improving the accuracy of sensitivity projections that considered only the target object in calculating signal power--in particular, the projections for dielectric haloscopes\cite{millar2017dielectric,brun2019new,baryakhtar2018axion,de2021dark}. The signal power depends on not only the dielectric stack, but also reflections from readout amplifiers/photon counters and preceding coupling elements, interactions of the dark-matter drive field directly with those elements, the electromagnetic properties of thermal and mechanical connections to the stack, and fields produced by nearby conductors (e.g., including coupling to the cavity modes of a shield/cryostat, similar to the dish). %Whereas analogous effects for receivers smaller than or comparable to the Compton wavelength are readily characterized (e.g., in lumped-element and cavity resonators by modifying a parameter in the equivalent circuit\cite{peng2000cryogenic,Chaudhuri:2014dla,chaudhuri2018fundamental}), in dielectric haloscopes, these effects give rise to interference, diffraction, and mode interactions that can be difficult to quantify and calibrate for accurate experimental sensitivity limits.%\footnote{For example, it has been observed that in a periodic transmission structure used for phase-coherent enhancement, a few percent of reflected power at the boundary suffices to produce a factor of two swing in receiver gain. Multiple reflection points produce further swings, as well as more complex frequency structure and ripples in the gain. See, e.g., Sec. 2.6 of \cite{pozar2012microwave}, \cite{eom2012wideband,chaudhuri2017broadband}, and references therein. In other words, a small error in the measurement of reflection can translate to a large uncertainty in dielectric haloscope boost factor.}
Indeed, the MADMAX collaboration has identified unwanted interference, as well as reflections from the horn antenna preceding the amplifier, as topics for investigation that can introduce significant difficulties and errors in calibration, substantially change the boost factor, and consequently, impact the ultimate sensitivity of its dielectric haloscope search\cite{egge2020first,knirck2021simulating}. The gain-bandwidth bound applies in the presence of these effects, making it a useful tool to validate or improve existing models\cite{millar2017dielectric,baryakhtar2018axion,knirck2019first}. Moreover, the gain-bandwidth bound applies in the presence of practical, complex electronic phenomena such as dielectric anisotropies, random and systematic variations in crystal properties across a disk and across layers (also identified as an important area for investigation by dielectric haloscope efforts), magneto-optical effects (e.g, Faraday rotation in a static magnetic field), and changes due to thermal and mechanical stresses (e.g., nonuniformity induced in propagation depths and resulting distortion in focal plane interference pattern, owing to warping at cryogenic temperature or under the influence of gravity)\cite{kittel1963quantum,ziman1972principles,beurthey2020madmax,krupka1999complex,makeev2002anisotropy,seshan2002handbook}; these effects on axion signal power and calibration are yet to be comprehensively studied. Our analysis can thus be used as a foundation for future investigations to produce more accurate sensitivity and calibration models, especially for multi-wavelength open receivers.
%, and to fully resolve the apparent tensions with planned experiments.

Finally, the constraints on impedance matching to axion dark matter are a critical input for search optimization, for which one must consider both signal and noise. The bounds on frequency-integrated signal power can be used to inform bounds on frequency-integrated signal-to-noise-ratio-squared, i.e. integrated sensitivity, which is the critical figure of merit for a search. Such an analysis is conducted in refs. \cite{chaudhuri2018fundamental,chaudhuri2019optimal}, which optimizes integrated sensitivity in single-moded, linear, passive receivers subject to the SQL on amplification. For more on the use of impedance matching constraints in refs. \cite{chaudhuri2018fundamental,chaudhuri2019optimal}, see Appendix \ref{ssec:SNR_opt}.

%------------------------------------------------------

\section{Conclusion}
\label{sec:conclude}

In this paper, we investigate impedance matching to axion dark matter, illustrating fundamental constraints on power absorption dictated by electromagnetic charges and their interaction with photons. We quantify backreaction on axion dark matter by calculating the axion source impedance. We consequently identify the self-impedance of the photon-electron interaction as the cause of large impedance mismatch with the axion source and poor power absorption. We formalize the limitation from the photon-electron interaction by deriving a gain-bandwidth bound on the match to axion dark matter, constraining power delivered to a linear, time-invariant, passive receiver integrated over a search band of possible rest-mass frequencies. We demonstrate the relationship between the derived gain-bandwidth bound and the Bode integral theorem and use the relationship to generate large classes of receivers, including the widely-used cavity resonator, that can saturate the bound. Our results, elucidating the role of the photon-electron interaction in laboratory searches, provide guidance for improvements in sensitivity. They illustrate apparent tension with the projections of a number of proposed experiments and provide a basis for more accurate sensitivity models, especially for multi-wavelength open receivers such as dish antennas and dielectric haloscopes. %They also yield constraints for signal-to-noise optimizations, such as refs. \cite{chaudhuri2018fundamental,chaudhuri2019optimal} for receivers subject to the SQL. 
\newline

The author is grateful to Kent Irwin, Lyman Page, Derek Jackson Kimball, Roman Kolevatov, Arran Phipps, and Nicholas Rapidis for comments on the manuscript. The author thanks Dick Bond, Peter Graham, Stephen Kuenstner, Betty Young, and the DM Radio Collaboration for enlightening discussions. The author is supported by the R.H. Dicke Postdoctoral Fellowship and the Wilkinson Research Fund at Princeton University. This research is funded in part by the Gordon and Betty Moore Foundation.

%-----------------------------------------------------

\appendix

\section{Impedance Matching For Background AC Fields and Hidden-Photon Dark Matter}
\label{sec:GBP_other}

We establish gain-bandwidth bounds on the impedance match for axions in a background, monochromatic AC field of finite spatial extent as well as for hidden-photon dark matter\cite{arias2012wispy,Chaudhuri:2014dla,Graham:2015rva}. Our treatment is analogous to that given in Section \ref{sec:GBP}.

\subsection{Background AC Field}
\label{ssec:GBP_AC}

A monochromatic AC magnetic field takes the form
\begin{equation}
    \vec{B}_{AC}(\vec{x},t)= \vec{B}^{(c)}(\vec{x})\textrm{cos}(\omega_{0}t) + \vec{B}^{(s)}(\vec{x})\textrm{sin}(\omega_{0}t).
\end{equation}
Such a background field is proposed for use in axion dark-matter searches based on upconversion.\cite{melissinos2008search,sikivie2010superconducting,Berlin:2019ahk} In these searches, the axion, of rest-mass frequency $\omega_{a}$, mixes with the AC magnetic field, producing electromagnetic fields and currents at the sideband frequencies $\omega_{0} \pm \omega_{a}$. One then aims to detect the signals at the sidebands.

We first derive the gain-bandwidth bound analogous to (\ref{eq:BF_DM}) and then discuss explicitly how it constrains power absorbed by a linear, time-invariant, passive receiver, integrated over all rest-mass frequencies in a search band. 

We assume that
\begin{equation}
    \vec{B}_{0}(\vec{x},t) \approx \vec{B}_{AC}(\vec{x},t),
\end{equation}
so that the effective axion current density is
\begin{equation}
    \vec{J}_{a}(\vec{x},t) \approx -\frac{\kappa_{a}}{Z_{fs}} \vec{B}_{AC}(\vec{x},t) \partial_{t}a(t).
\end{equation}
In analogy with eq. (\ref{eq:IVat}), we define two pairs of quantities that play the role of current and voltage. For the cosine component of the background field, we define
\begin{equation}\label{eq:IVat_c}
    I_{a}^{(c)}(t) \equiv \frac{\kappa_{a}}{Z_{fs}}\mathrm{cos}(\omega_{0}t)\partial_{t}a,\quad V_{a}^{(c)}(t) \equiv \int_{V} \vec{E}(\vec{x},t) \cdot \vec{B}^{(c)}(\vec{x})
\end{equation}
and similarly $I_{a}^{(s)}(t)$ and $V_{a}^{(s)}(t)$ for the sine component.

%\begin{equation}\label{eq:IVat_s}
%    I_{a}^{(s)}(t) \equiv \frac{\kappa_{a}}{Z_{fs}}\mathrm{sin}(\omega_{0}t)\partial_{t}a,\ V_{a}^{(s)}(t) \equiv \int_{V} \vec{E}(\vec{x},t) \cdot \vec{B}^{(s)}(\vec{x})
%\end{equation}

Conservation of energy dictates (cf. (\ref{eq:axion_Poynting}))
\begin{equation}
    I_{a}^{(c)}(t)V_{a}^{(c)}(t) + I_{a}^{(s)}(t)V_{a}^{(s)}(t) = P_{J}(t) + P_{rad}(t) + \partial_{t}(W_{B}(t)+W_{E}(t)). 
\end{equation}
We assume that the relationship between $I_{a}^{(c)}(t)$ and $V_{a}^{(c)}(t)$ is linear and time-invariant (and identically for $I_{a}^{(s)}(t)$ and $V_{a}^{(s)}(t)$). Then, we may define impedance functions $Z^{(c)}(t)$ and $Z^{(s)}(t)$ which satisfy (cf. (\ref{eq:Zdef}))
\begin{equation}\label{eq:Zdef_cs}
    V_{a}^{(c,s)}(t)=\int_{-\infty}^{\infty}\ d\tau\ Z_{a}^{(c,s)}(t-\tau)I_{a}^{(c,s)}(\tau).
\end{equation}
%\begin{equation}\label{eq:Zdef_s}
%    V_{a}^{(s)}(t)=\int_{-\infty}^{\infty}\ d\tau\ Z_{a}^{(s)}(t-\tau)I_{a}^{(s)}(\tau).
%\end{equation}

From impulse-response arguments, we find (cf. (\ref{eq:BF_DM}))
\begin{equation}\label{eq:BF_DM_cs}
    \int_{0}^{\infty}d\omega\ R_{a}^{(c,s)}(\omega) \leq \frac{\pi}{2\epsilon_{0}} \int_{V} |\vec{B}^{(c,s)}(\vec{x})|^{2}.
\end{equation}
%\begin{equation}\label{eq:BF_DM_s}
%    \int_{0}^{\infty}d\omega\ R_{a}^{(s)}(\omega) \leq \frac{\pi}{2\epsilon_{0}} \int_{V} |\vec{B}^{(s)}(\vec{x})|^{2}
%\end{equation}
$R_{a}^{(c)}(\omega)$ and $R_{a}^{(s)}(\omega)$ are the real parts of the Fourier transforms of $Z^{(c)}(t)$ and $Z^{(s)}(t)$. The volume $V$ contains the background field and all axion-driven currents.

To demonstrate how eq. (\ref{eq:BF_DM_cs}) constrains the power delivered to a receiver, we consider a situation resembling a laboratory search. (See Section \ref{sec:GBP}.) We calculate an upper bound on the receiver power dissipation, integrated over a search band.
%We assume, as is proposed for upconversion schemes, that these frequencies are less than the frequency of the background field, $\omega_{l},\omega_{h} < \omega_{0}$.
An axion field oscillating at rest-mass frequency $\omega_{a}$, as in eq. (\ref{eq:axion_field}), gives rise to Fourier components in the currents $I_{a}^{(c)}(t)$ and $I_{a}^{(s)}(t)$ at the sideband frequencies $\omega_{0} \pm \omega_{a}$. 
%The magnitude of these Fourier components is given by
%\begin{equation}
%    |I_{a}^{(c)}(\omega_{0} \pm \omega_{a})|^{2} = |I_{a}^{(s)}(\omega_{0} \pm \omega_{a})|^{2}= \frac{1}{8} \frac{(\kappa_{a} c)^{2}}{\mu_{0}} \rho_{a}.
%\end{equation}
Analogous to eq. (\ref{eq:Pdiss_FFT}),
\begin{equation}\label{eq:Pdiss_FFT_AC}
    P(\omega_{a}) = \frac{1}{4} \frac{(\kappa_{a} c)^{2}}{\mu_{0}} \rho_{a} (R_{a}^{(c)}(\omega_{0}+ \omega_{a}) + R_{a}^{(c)}(\omega_{0} - \omega_{a}) + R_{a}^{(s)}(\omega_{0} + \omega_{a}) +R_{a}^{(s)}(\omega_{0} - \omega_{a}) ) 
\end{equation}
represents the power flow out of the axion field. The power dissipated in the charges is at most the power flow out of the axion field, so integrating (\ref{eq:Pdiss_FFT_AC}) yields
\begin{equation}\label{eq:PJ_DM_AC}
     \int_{\omega_{l}}^{\omega_{h}} d\omega_{a}\ P_{J}(\omega_{a})  \leq   \frac{1}{2} \frac{\pi}{2} (\kappa_{a}c^{2})^{2} \rho_{a} \int_{V} (|\vec{B}_{AC}^{(c)}(\vec{x})|^{2}+ |\vec{B}_{AC}^{(s)}(\vec{x})|^{2}).  
\end{equation}
Here, we have used (\ref{eq:BF_DM_cs}) and the fact that $R_{a}^{(c)}(\omega)$, $R_{a}^{(s)}(\omega)$ are even functions of frequency. %Note that (\ref{eq:PJ_DM_AC}) is nearly identical to (\ref{eq:PJ_DM}), with the intensity of the DC magnetic field replaced with the time-averaged intensity of the AC magnetic field.

\subsection{Hidden-Photon Dark Matter}
\label{ssec:GBP_HP}

The hidden photon is a vector particle, characterized by mass $m_{\gamma'}$ and three-vector potential $\vec{A}'(\vec{x},t)$. Like the axion, the effect of the hidden photon on Maxwell's equations is to add a current density\cite{Graham:2014sha,Chaudhuri:2014dla}: denoting $\varepsilon$ as the kinetic mixing angle,
%\begin{equation}\label{eq:rho_HP}
%    \rho_{\gamma'}(\vec{x},t)=-\varepsilon \frac{1}{\mu_{0}} \left( \frac{m_{\gamma'}c^{2}}{\hbar} \right)^{2} \phi'(\vec{x},t)
%\end{equation}
\begin{equation}\label{eq:J_HP}
    \vec{J}_{\gamma'}(\vec{x},t)=-\varepsilon \epsilon_{0} \left( \frac{m_{\gamma'}c^{2}}{\hbar} \right)^{2} \vec{A}'(\vec{x},t).
\end{equation}
Unlike the axion, the effective hidden-photon current density fills all space.

While the impedance-matching constraint (\ref{eq:BF_DM}) for axions is set by the volume of the DC magnetic field, the constraint for hidden photons is set by the volume $V_{s}$ of the high-conductivity shield surrounding the receiver, which is necessary to mitigate electromagnetic interference in a search. For a high-conductivity (superconducting) shield which is many skin (penetration) depths thick at all frequencies within the search range, the receiver couples negligible power from the hidden-photon field outside of the shield.\cite{Graham:2014sha}  We assume that the hidden photon field is spatially uniform within the shielded volume, which is appropriate when the coherence length is much larger than the length scale of the shield. Suppose the field points in the $\hat{z}$ direction. Then, the role of current and voltage are played by
\begin{equation}
    I_{\gamma'}(t) \equiv -\varepsilon \epsilon_{0} \left( \frac{m_{\gamma'}c^{2}}{\hbar} \right)^{2} A'_{z}(t),\ V_{\gamma'}(t) \equiv \int_{V_{s}} E_{z}(\vec{x},t).
\end{equation}
The subscript $z$ represents the $z$-component of the vector. 

For a linear, time-invariant, passive receiver, we define an impedance function given by
\begin{equation}
    V_{\gamma'}(t)=\int_{-\infty}^{\infty}\ d\tau\ Z_{\gamma'}(t-\tau)I_{\gamma'}(\tau).
\end{equation}
Denote $R_{\gamma'}(\omega)$ as the real part of the Fourier transform of $Z_{\gamma'}(t)$. Using impulse response arguments, we find
\begin{equation}\label{eq:BF_HP}
    \int_{0}^{\infty} d\omega\ R_{\gamma'}(\omega) \leq \frac{\pi}{2} V_{s}
\end{equation}
A similar argument is used for shielded volumes in Section \ref{ssec:impulse_correct}.
%In establishing (\ref{eq:BF_HP}), we have extrapolated the boundary conditions describing the shield. See also Section \ref{ssec:impulse_correct}.

%To see how (\ref{eq:BF_HP}) constrains the power absorbed in a receiver, integrated across a search band, we consider a situation resembling a laboratory search for hidden-photon dark-matter. Consider a spatially uniform hidden-photon field of mass $m_{\gamma'}$, oscillating monochromatically at frequency $\omega_{\gamma'}=m_{\gamma'}c^{2}/\hbar$. Assume that it points in the $\hat{z}$ direction. We may then write the hidden-photon vector potential and energy density as
%\begin{equation}
%    \vec{A}'(t)=Re(\tilde{A}' \textrm{exp}(+i\omega_{\gamma'}t)) \hat{z}
%\end{equation}
%\begin{equation}
%    \rho_{\gamma'}= \frac{\epsilon_{0}}{2} \omega_{\gamma'}^{2} |\tilde{A}'|^{2}.
%\end{equation}
Let $\omega_{\gamma'}=m_{\gamma'}c^{2}/\hbar$.
We hold fixed the hidden-photon mixing angle $\varepsilon$ and energy density $\rho_{\gamma'}$ across a search band. Similar to Section \ref{sec:GBP}, we find that the receiver power dissipation must satisfy
\begin{equation}\label{eq:PJ_HP}
    \int_{\omega_{l}}^{\omega_{h}} \frac{d\omega_{\gamma'}}{\omega_{\gamma'}^{2}} P_{J}(\omega_{\gamma'}) \leq \frac{\pi}{2} \varepsilon^{2} \rho_{\gamma'} V_{s}.
\end{equation}
For axions in background AC fields and hidden photons, we may use the methods of Section \ref{sec:search_opt} to determine receivers saturating the gain-bandwidth bounds. One may note that the limitations on impedance matching observed in Section \ref{sec:gedank} are qualitatively the same for axions in background AC fields and hidden photons.

%------------------------------------------------------

\section{Previous Work}
\label{sec:other_work}

\subsection{Signal-to-Noise Optimization}
\label{ssec:SNR_opt}

We place our impedance-matching analysis in the context of the optimization undertaken in refs. \cite{chaudhuri2018fundamental,chaudhuri2019optimal} for single-moded, linear, passive receivers subject to the SQL on phase-insensitive amplification. While the match to axion dark matter is a necessary component of search optimization, it is certainly not sufficient to consider only signal power. Optimizing search sensitivity requires a simultaneous consideration of signal and measurement noise, necessitating a far more comprehensive analysis.

One must study how dark-matter power is coupled into the receiver charges and transferred to the readout. As described in Sections II and III of \cite{chaudhuri2018fundamental}, the conceptual receivers of our thought experiment represent the two generic categories of coupling to the axion-induced electromagnetic fields, radiative coupling and reactive coupling. For readout with a phase-insensitive amplifier subject to the SQL, receivers using solely radiative couplings are generally disadvantaged in scan sensitivity, compared to reactive couplings, because of the mismatch between the effective dark-matter source impedance and the free-space self-impedance. This motivates a focus on single-moded, reactively coupled receivers, which describes the majority of electromagnetic axion dark-matter receivers\cite{sikivie2021invisible}. These receivers are subject to the voltage constraint (\ref{eq:VL_ineq}), derived from the gain-bandwidth bound.

A receiver optimization must, at minimum, consider thermal noise and readout noise. For readout with a phase-insensitive amplifier, the frequency response of a receiver to backaction-noise forcing and to axion forcing (e.g., $\tilde{V}_{L}(\omega_{a})$ in Fig. \ref{fig:IndBF}) are generally not the same.
One must then produce a framework for analyzing signal and noise transfer throughout a receiver, considering simultaneously power matching and amplifier-noise matching. Thus, the critical figure of merit is not frequency-integrated signal power, but frequency-integrated signal-to-noise-ratio-squared, i.e. integrated sensitivity. To maximize integrated sensitivity, one would ideally achieve an efficient power match to dark matter and an efficient noise match to the amplifier at all frequencies. However, the ability to do so is severely limited for a reactively coupled receiver read out by an amplifier subject to the SQL. In ref. \cite{chaudhuri2018fundamental}, the metric of integrated sensitivity is constrained with the Bode-Fano criterion, an extension of the Bode integral theorem\cite{bode1945network,fano1950theoretical}. Note that while the Bode-Fano criterion is typically used to constrain gain-bandwidth product on signal transfer (similar to Section \ref{ssec:Bode}), ref. \cite{chaudhuri2018fundamental} uses it to constrain gain-bandwidth product on signal-to-noise. An optimized single-pole cavity resonator can achieve an integrated sensitivity that is approximately 75\% of the Bode-Fano limit, while a multi-pole Chebyshev filter can saturate the limit. Additionally, while high-$Q$ resonant and broadband search strategies possess comparable integrated signal power, the difference in integrated sensitivity is orders of magnitude. Saturating the limit on integrated sensitivity is thus distinct from saturating the limit on integrated signal power.

In this paper, we considered the impedance match of dark matter to a single receiver with parameters that are constant in time. A single-pole resonator, while possessing near-ideal integrated sensitivity, has poor sensitivity far off-resonance. One needs to tune the resonant frequency to conduct a sensitive probe over a wide search range. An optimization of scan strategy must then account for periodically-varied frequency response. We must also incorporate prior information on the axion signal. These priors may take the form of previous constraints, as well as regions of parameter space that are particularly well-motivated, e.g., the QCD axion band\cite{graham2015experimental}. One can calculate without a detailed consideration of priors that, subject to the SQL, a resonant search is superior to a reactive broadband search at all frequencies at which a resonator may practically be constructed. However, priors are required to calculate the size of the advantage and to conclude that, for most reasonable priors over a wide search range, the advantage is a few orders of magnitude in integration time.

Altogether, a large number of factors---the match to dark matter, signal and noise transfer, thermal and readout noise (including backaction), noise matching, periodically-varied frequency-response, and priors---must be considered simultaneously for a comprehensive search optimization, as analyzed in refs. \cite{chaudhuri2018fundamental,chaudhuri2019optimal} for single-moded linear, passive receivers, subject to the SQL.

\subsection{Impulse Response}
\label{ssec:imp_response}

Before discussing ref. \cite{lasenby2021parametrics}, it is first useful to demonstrate that (\ref{eq:BF_DM}) holds with equality for a system without charges. 
%We determine $R_{a}(\omega)$ using eq. (\ref{eq:Rdef}). 
%We calculate the complex-valued electric $\vec{E}(\vec{x},\omega)$ and magnetic $\vec{B}(\vec{x},\omega)$ fields in the far field of the DC magnetic field, 
Given $I_{a}(\omega)$, the complex-valued vector potential $\vec{A}(\vec{x},\omega)$ is, in the far-field limit $(\omega/c)|\vec{x}| \gg 1$,\cite{Jackson:1998nia}
\begin{equation} \label{eq:mag_farfield}
    \vec{A}(\vec{x},\omega) \rightarrow \frac{\mu_{0}}{4\pi} I_{a}(\omega) \frac{\mathrm{exp}(-ik|\vec{x}|)}{|\vec{x}|} \int_{V} d^{3}\vec{x}' \vec{B}_{DC}(\vec{x}') \mathrm{exp}(+ik \hat{x} \cdot \vec{x}'), 
\end{equation}
where the volume $V$ entirely surrounds the DC magnetic field. $\hat{x}$ is a unit vector pointing from the origin to the position $\vec{x}$, and $k=\omega/c$. %The electric and magnetic fields in the far field are given by
%\begin{equation}
%    \vec{E}(\vec{x},\omega)=-i\omega \vec{A}(\vec{x},\omega)
%\end{equation}
%\begin{equation}
%    \vec{B}(\vec{x},\omega)=-ik\hat{x} \times \vec{A}(\vec{x},\omega)
%\end{equation}
Consider a sphere $S$, centered at the origin, with all points on the surface in the far field. Eq. (\ref{eq:Rdef}) yields (see Ch. 9.1 of ref. \cite{Jackson:1998nia})
\begin{align}
    R_{a}(\omega) &= \frac{1}{\mu_{0}|I_{a}(\omega)|^{2}} \int_{\vec{x} \in S} da\  (\vec{E}(\vec{x},\omega) \times \vec{B}^{*}(\vec{x},\omega))\cdot \hat{x} \nonumber \\
    &= \frac{1}{Z_{fs}|I_{a}(\omega)|^{2}}  \int_{\vec{x} \in S} da\ \omega^{2} |\vec{A}(\vec{x},\omega)|^{2} %\\
%    &=\frac{\omega^{2} \mu_{0}^{2}}{16 \pi^{2} Z_{fs}} \int d\Omega \left| \int_{V} d^{3}\vec{x}' \vec{B}_{DC}(\vec{x}') \mathrm{exp}(+ik \hat{x} \cdot \vec{x}') \right|^{2} \nonumber 
\end{align}
Plugging in eq. (\ref{eq:mag_farfield}) and simplifying yields
\begin{equation}\label{eq:Raw}
    R_{a}(\omega) = \frac{\omega^{2} \mu_{0}}{4\pi c} \int_{V} d^{3}\vec{x}' \int_{V} d^{3}\vec{x}'' \vec{B}_{DC}(\vec{x}') \cdot \vec{B}_{DC}(\vec{x}'') \mathrm{sinc}(k|\vec{x}'-\vec{x}''|)
\end{equation}
To integrate this expression over all frequencies, note that
%\begin{equation}\label{eq:Dirac}
%    \delta(\vec{x}'-\vec{x''}) = \frac{1}{2\pi^{2}} \int_{0}^{\infty} dk\ k^{2} \mathrm{sinc}(k|\vec{x}'-\vec{x}''|),
%\end{equation}
\begin{align}
    \delta(\vec{x}'-\vec{x''}) &= \frac{1}{(2\pi)^{3}} \int d^{3} \vec{k}\ \mathrm{exp}(+i \vec{k} \cdot (\vec{x}'-\vec{x}'')) \nonumber \\
    &= \frac{1}{2\pi^{2}} \int_{0}^{\infty} dk\ k^{2} \mathrm{sinc}(k|\vec{x}'-\vec{x}''|),\label{eq:Dirac}
\end{align}
where the first integral is over all real-valued vectors $\vec{k}$. 
Combining with (\ref{eq:Raw}) shows that (\ref{eq:BF_DM}) holds with equality.

Ref. \cite{lasenby2021parametrics} introduces the impulse-response approach to constrain power flow from the axion field and derives a generic gain-bandwidth relationship on the impedance match to axion dark matter which is similar to eq. (\ref{eq:BF_DM}). (See also ref. \cite{baryakhtar2018axion}, which uses the impulse-response approach to specifically calculate frequency-averaged signal power in an axion haloscope consisting of stacks of dielectrics.) However, the relationship is an equality, rather than the inequality derived in our paper. We now detail how the work in our paper differs from and extends \cite{lasenby2021parametrics}. Most importantly, our mathematical treatment directly includes electrons and thus can be applied to existing laboratory searches probing the axion-photon coupling. We summarize arguments in Sections IIA and IIB of \cite{lasenby2021parametrics}. 

Ref. \cite{lasenby2021parametrics} factorizes the background field (discussion following (14) of \cite{lasenby2021parametrics}), $B_{0}(x,t)=B_{0}(t)b(x)$. We focus on DC magnetic fields, for which $B_{0}(t)=B_{0}$ is time-independent. Ref. \cite{lasenby2021parametrics} defines
\begin{equation}\label{eq:EbAb_def}
    E_{b}(t) \equiv \frac{1}{V_{b}} \int E \cdot b(x),\quad A_{b}(t) \equiv \frac{1}{V_{b}} \int A \cdot b(x)
\end{equation}
where $V_{b}=\int b^{2}(x)$ and the integrals are performed over all space. $E_{b}(t)$ is proportional to $V_{a}(t)$, as defined in eq. (\ref{eq:IVat}). $A=A(x,t)$ is the vector potential in Coulomb gauge. Note that $\dot{A}_{b}(t)=-E_{b}(t)$. $E_{b}(t)$ obeys
\begin{equation}\label{eq:Eb_EOM}
    \dot{E}_{b}(t) = \frac{1}{V_{b}} \Bigg( -\int d^{3}x\ \nabla^{2}A(x,t) \cdot b(x) - g j(t) - \int d^{3}x\ J(x,t) \cdot b(x) \Bigg) 
\end{equation}
where $g$ is the axion-photon coupling, $j(t)=\dot{a}(t) B_{0} V_{b}$ is proportional to the effective axion current, and $J(x,t)$ represents electromagnetic current. For consistency with ref. \cite{lasenby2021parametrics}, we use natural units $\hbar=c=\epsilon_{0}=1$ here.

Ref. \cite{lasenby2021parametrics} considers the response of the variable $E_{b}(t)$ to a short $j(t)$ pulse. See the paragraph following eq. (17) of ref. \cite{lasenby2021parametrics}. This pulse effectively takes the form $j(t)=J\delta(t)$. Assume $E_{b}(t)$ is zero before the pulse. Using (\ref{eq:Eb_EOM}), the change in $E_{b}(t)$ due to the pulse is
\begin{equation}\label{eq:Eb_change}
    \Delta E_{b}= -\frac{gJ}{V_{b}} -\frac{1}{V_{b}} \lim_{\tau \to 0+}  \int_{-\infty}^{\tau} dt \Bigg( \int d^{3}x\ \nabla^{2}A(x,t) \cdot b(x) + \int d^{3}x\ J(x,t) \cdot b(x) \Bigg) 
\end{equation}
$\Delta E_{b}$ is the analog of $\int \vec{E}^{\delta}(\vec{x},0+) \cdot \vec{B}_{DC}(\vec{x})$ in our main-text analysis. In assuming that the pulse occurs ``much faster than the system's dynamics,'' ref. \cite{lasenby2021parametrics} excludes the last two terms on the right-hand side of eq. (\ref{eq:Eb_change}), which incorporate the contribution of charges to the electric-field response. The energy from the pulse is then
\begin{equation}\label{eq:work_wrong}
    <W>=\frac{1}{2}(\Delta E_{b})^{2} V_{b} = \frac{g^{2}J^{2}}{2 V_{b}}.
\end{equation}
Using this equation, along with methods similar to eq. (\ref{eq:work_GBP}), ref. \cite{lasenby2021parametrics} arrives at
\begin{equation}\label{eq:sum_rule}
    \int_{-\infty}^{\infty} d\omega\ \omega \textrm{Im} \tilde{\chi}(\omega) = \frac{\pi}{V_{b}},
\end{equation}
describing the gain-bandwidth product of the impedance match to axion dark matter (see eq. (20) of \cite{lasenby2021parametrics}). $\tilde{\chi}(\omega)$ is the linear-response function describing how $A_{b}$ responds to the axion forcing. This result is similar to (\ref{eq:BF_DM}), except that it is an equality.

Note that by excluding the last two terms on the right-hand side of eq. (\ref{eq:Eb_change}), the derivation in \cite{lasenby2021parametrics} is independent of the dynamics of receiver electrons. In contrast, our work accounts for the effect of the electrons on the impedance match to axions. We have shown in the main text that once electrons are explicitly included and the assumptions for the gain-bandwidth argument are defined (in particular, the assumption of passivity (\ref{eq:EM_passive})), then one obtains inequality instead of equality. We have described how the resulting bounds on frequency-integrated absorbed power (see eq. (\ref{eq:PJ_DM})) can be evaded with active matching elements, e.g. negative inductors enabling wideband reactance cancellation\cite{sussman2009non,salit2010enhancement,shlivinski2018beyond}.

Note that the assumptions of ref. \cite{lasenby2021parametrics} are fulfilled in the absence of charges. In this case, axion-field power is converted solely into electromagnetic radiation. As calculated above using eqs. (\ref{eq:Raw})- (\ref{eq:Dirac}), one then finds that (\ref{eq:BF_DM}) holds with equality, consistent with \cite{lasenby2021parametrics}. 

In summary, the impulse response approach and the frontier (\ref{eq:sum_rule}) proposed in ref. \cite{lasenby2021parametrics} are a significant step in understanding first-principles constraints on coupling power from the axion field, integrated over possible rest-mass frequencies in a search range. In this paper, we have extended the impulse response approach to incorporate the effect of electrons and the photon-electron interaction in coupling power from the axion field. We have shown stringent constraints on the impedance match to axion dark matter for linear, time-invariant, passive receivers. We have explained why gain-bandwidth bounds on signal power are necessary, but not sufficient, to set gain-bandwidth bounds on signal-to-noise ratio or constrain search sensitivity. Constraining and optimizing signal-to-noise ratio requires a far more comprehensive analysis, as performed in refs. \cite{chaudhuri2018fundamental,chaudhuri2019optimal} for single-moded, linear, passive receivers subject to the SQL. Importantly, we have shown apparent tension between the gain-bandwidth bound on signal power and the sensitivity projections of a number of planned experiments. We have provided a foundation for more accurate signal-power calculations, especially for multi-wavelength open receivers such as dish antennas and dielectric haloscopes.

\bibliographystyle{JHEP}
\bibliography{Bib_Match}

\end{document}